\documentclass[10pt,journal]{IEEEtran}
\IEEEoverridecommandlockouts
\usepackage[dvips]{graphicx}
\usepackage[colorlinks,urlcolor=blue,linkcolor=blue,citecolor=blue]{hyperref}
\usepackage{color,array}
\usepackage{xspace}
\usepackage{amsmath}
\usepackage{caption}
\usepackage{subcaption}
\usepackage{placeins}
\usepackage{balance}
\usepackage{cite}
\usepackage{bm}
\usepackage{algpseudocode}
\usepackage{amsmath}
\usepackage{array}
\usepackage{setspace} 
\usepackage[nameinlink]{cleveref}
\usepackage{accents}
\usepackage{mathtools}
\usepackage[dvipsnames]{xcolor}
\usepackage{amsthm} 
\usepackage{multirow}
\usepackage[linesnumbered,ruled,vlined]{algorithm2e}
\usepackage{tabularx}
\SetKwRepeat{Do}{do}{while}
\usepackage{color}
\usepackage{xcolor}
\usepackage{epstopdf}
\usepackage{endnotes}
\usepackage{framed}
\usepackage{float}
\usepackage{cool} 
\usepackage{enumerate} 
\usepackage{pifont}
\Crefname{figure}{Fig.}{Figs.}

\usepackage{mathrsfs}
\usepackage[justification=centering]{caption}
\allowdisplaybreaks[4]

\newcolumntype{P}[1]{>{\centering\arraybackslash}p{#1}}

\DeclareUnicodeCharacter{2212}{-}

\usepackage{changes}
\usepackage{lipsum}
\definechangesauthor[name={Moqbel1}, color=red]{Revised/Added}
\definechangesauthor[name={Moqbel2}, color=red]{Deleted}
\definechangesauthor[name={Moqbel3}, color=red] {moved to appendix E}

\usepackage{color}
\usepackage{epstopdf}
\def\BibTeX{{\rm B\kern-.05em{\sc i\kern-.025em b}\kern-.08em
        T\kern-.1667em\lower.7ex\hbox{E}\kern-.125emX}}
\title{Efficient Data Labeling and Optimal Device Scheduling in HWNs Using Clustered Federated Semi-Supervised Learning}
\author{Moqbel Hamood,
      Abdullatif Albaseer,~\IEEEmembership{Member,~IEEE,}
        Mohamed Abdallah,~\IEEEmembership{Senior Member,~IEEE,}
        Ala Al-Fuqaha,~\IEEEmembership{Senior Member,~IEEE}
    \thanks{Moqbel Hamood, Abdullatif Albaseer, Mohamed Abdallah and~Ala Al-Fuqaha and are with the Division of Information and Computing Technology, College of Science and Engineering, Hamad Bin Khalifa University, Doha, Qatar (e-mail:\{moha19838,aalbaseer, moabdallah, aalfuqaha\}@hbku.edu.qa).}
    
 \thanks{* Preliminary results in this work are accepted at the IEEE WCNC Conference, 2024~\cite{hamood2024empowering}.
}
}
\begin{document}
\maketitle
\begin{abstract}
Clustered Federated Multi-task Learning (CFL) has emerged as a promising technique to address statistical challenges, particularly with non-independent and identically distributed (non-IID) data across users. However, existing CFL studies entirely rely on the impractical assumption that devices possess access to accurate ground-truth labels. This assumption becomes specifically problematic in hierarchical wireless networks (HWNs), with vast unlabeled data and dual-level model aggregation, not only leading to slowing down convergence speeds and extending processing times but also resulting in increased resource consumption. To this end, we propose Clustered Federated Semi-Supervised Learning (CFSL), a novel framework tailored for more realistic scenarios in HWNs. We leverage specialized models resulting from device clustering and present two prediction model schemes, the best-performing specialized model and the weighted-averaging ensemble model, to correctly label unlabeled, unseen data. For the best-performing specialized model scheme, a specialized model excelling in label prediction for a specific device is assigned to correctly label the unlabeled data, even when the data originates from other environments, while the weighted-averaging ensemble model combines all specialized models into a unified model, capturing more details from broader data distributions across edge networks. The CFSL also introduces two novel prediction time schemes, split-based and stopping-based, for accurately timing the labeling process, alongside two strategic device selection schemes, greedy and round-robin, upon reaching each cluster's stopping point. Extensive testing validates CFSL’s superiority over existing models in labeling and testing accuracies and resource efficiency, achieving up to 51$\%$ energy savings.
\end{abstract}
\section{Introduction}
\IEEEPARstart{I}n recent years, the rapid proliferation of mobile and Internet of Things (IoT) devices, propelled by advancements in beyond-fifth-generation (B5G) technologies, has sparked extensive data generation that has significantly contributed to understanding the current system behavior or predicting future states. This data generation not only transforms everyday life with multifaceted applications such as human activity recognition~\cite{ chen2012sensor,ramanujam2021human} and computer vision~\cite{voulodimos2018deep,han2022survey} but also enhances inter-domain knowledge sharing across various edge networks within Hierarchical Wireless Networks (HWNs). This leads to unprecedented collaboration in these networks~\cite{sun2019application,hosseinalipour2020federated} with significant benefits such as seamless data transmission within different tiers, enhancing communication reliability, and optimizing data handling efficiency across the network hierarchy. However, the swift increase in data generated by devices (i.e., mobile phones and intelligent vehicles) in various applications poses significant challenges. These applications highly depend on supervised learning, requiring high-quality labeled data for training classifiers. Furthermore, offloading this data to edge and cloud servers for machine learning analysis is costly, time-consuming, and fraught with privacy and complexity issues. Additionally, the different processing capabilities of devices make it difficult to conduct training directly on user devices, especially for computation-intensive applications. These challenges complicate utilizing large datasets for predictive analytics on each device.

In tackling the abovementioned challenges, Hierarchical Federated Learning (HFL)~\cite{konevcny2016federated,abad2020hierarchical,chen2022federated,albaseer2021fine,luo2020hfel} has recently emerged as a promising solution, attracting significant interest from researchers since it keeps data at its original location, protecting privacy. Devices in this approach play an active role in model training by sending only model parameters (i.e., weights and biases) to edge servers (i.e., updating edge models) and cloud (i.e., updating the global model)~\cite{xu2021federated}. After that, the global model is sent to edge servers and workers for further training. This process continues iteratively until the global model converges to the optimal solution. However, HFL faces critical challenges, primarily due to the unrealistic assumption that edge devices have sufficient labeled data for training. This is violated in practical scenarios where a substantial portion of the data on these devices is unlabeled. Besides, challenges are compounded by unbalanced and non-independent and identically distributed (non-IID) data and tremendous devices with different capabilities~\cite{li2021federated}. These challenges collectively prevent efficient model communication across the wireless channels of HWNs.

In response, Clustered Federated Learning (CFL) was introduced as an appealing approach specifically designed to deal with statistical heterogeneity challenges~\cite{sattler2020clustered,mansour2020three,ghosh2020efficient,albaseer2021client,luo2021energy,ma2022convergence}. The CFL exploits the geometric relationships between workers exhibiting similar data generation patterns. This approach involves developing a clustering mechanism that groups workers based on the similarity of their data distributions, with each cluster assigned a specialized model aligned to its specific data distribution~\cite{kim2021dynamic,gong2022adaptive,hamood2023intelligent}. By clustering similar data patterns and facilitating knowledge sharing within each cluster, we enable multi-task learning through the creation of models tailored to the unique needs and distributions of each cluster. Here, CFL initiates the clustering process once the HFL objective's solution reaches a stationary point (i.e., where the global model cannot improve further on its current data). Thus, CFL considers a post-processing technique to improve the learning performance of the HFL without altering its core communication protocols. Extensive research has been dedicated to the CFL using labeled data in wireless edge networks, addressing critical issues such as learning performance, bandwidth scarcity, and energy consumption~\cite{huang2023active,albaseer2023fair,hamood2023clustered}. Recent studies have focused on CFL's effectiveness and proposed novel solutions to support its applications. For instance, authors in~\cite{huang2023active} investigated the worker selection mechanism by selecting the most informative workers to aggregate their models and update the global model. The study in~\cite{albaseer2023fair} has optimized worker participation based on training latency, exploiting the dynamic nature of wireless edge networks to schedule devices based on their latency metrics. Despite these advancements, these studies have yet to address the abundance of unlabeled data among workers, neither in edge nor hierarchical networks.  In addition, selecting only a limited number of workers for training forces the cloud and edge servers in HWNs to follow strict worker scheduling strategies.

Semi-Supervised Learning (SSL) has emerged as an attractive strategy for addressing the challenges of insufficient labeled data by leveraging small amounts of labeled data to label larger volumes of unlabeled data. In the literature, several studies have made considerable efforts to apply the SSL in conjunction with FL to enhance model performance and learning efficiency in wireless networks~\cite{albaseer2022semi,diao2022semifl,presotto2022semi,10159650,tashakori2023semipfl}. Diao et al.~\cite{diao2022semifl} proposed an approach incorporating SSL and FL, in which workers held unlabeled data and servers contained the labeled data to strengthen the learning process. Albaseer et al.~\cite{albaseer2022semi} explored adapting SSL within FL, considering bandwidth, latency, and energy consumption constraints in wireless edge networks. Despite their efforts,  none of these studies have considered practical scenarios where most data is unlabeled in unbalanced and non-IID and device incompatibility settings. These studies also ignore more realistic scenarios when dealing with unlabeled data in HWNs, where multiple edge networks seek to collaborate and capture different patterns from these networks, making Problems more complicated.

In practice, the majority of data collected by edge devices in HWNs is unlabeled, necessitating advanced solutions that leverage this unlabeled data in non-IID data and limited resources. A promising approach is the integration of SSL with CFL in HWN settings to tackle these challenges effectively. Since CFL develops multiple specialized models tailored to specific device groups, using these models with SSL to label unlabeled data, mainly when the data originates from different distributions, presents significant challenges. The complexity of this task is further complicated by the multiple layers and varied computational processes inherent in devices within HWNs, leading to increased resource consumption. Therefore, it is essential to develop a novel framework that leverages SSL and CFL to optimize unlabeled data across edge networks in HWNs while simultaneously addressing the problems of non-IID data distribution and resource constraints.

Spurred by these unmet challenges, we propose a novel framework that integrates CFL with SSL in HWNs, aiming to fill the existing research gaps. This framework, dubbed Clustered Federated Semi-Supervised Learning (CFSL), is tailored for more practical scenarios where a large portion of the data is unlabeled and a minimal subset of the data is labeled in workers under non-IID data settings within the same or even across different edge networks. Specifically, the proposed solution comprises two prediction model schemes, two prediction time schemes, and two worker selection strategies. Our main contributions are listed as follows.
\begin{itemize}
   \item Proposing a novel framework, CFSL, implemented in HWNs to address the challenges of scarce labeled data in real-world applications while optimizing resource consumption. This reduces training time and energy consumption while effectively handling unbalanced and non-IID data through hierarchical clustering and adequate labeling.
    \item Formulating a joint optimization problem for labeling unlabeled data in HWNs by minimizing the loss function and maximizing the model utility under system constraints, leading to improving model performance while minimizing training time and resource costs. Since the problem is intractable, we introduce a heuristic-based solution leveraging the abundance of unlabeled data in non-IID settings and system heterogeneity, providing practical solutions.
  \item Conducting experimental analysis using realistic datasets (i.e., FEMNIST and CIFAR-10)in unbalanced and non-IID settings. The experiment results demonstrate the proposed approach's superiority over baselines in enhancing learning performance while reducing training time and resource consumption.
\end{itemize}
This paper is organized as follows: Section~\ref{related_work} reviews the related work. In Section \ref{Sys_model}, we elaborate on the system model, including the learning process, and outline the computation and communication models. Section \ref{problem_form} is dedicated to formulating the optimization problem. Our proposed solution is detailed in Section \ref{proposed_sol}. Convergence analysis is introduced in Section~\ref{Conv}. Experimental results and findings are presented in Section \ref{results}. The paper concludes with Section \ref{conclusion} and suggests directions for future investigations.
\section{Related Work}
\label{related_work}
HFL within HWNs has emerged as a significant area of research in recent years. Numerous studies have focused on addressing the limitations of traditional wireless edge networks by leveraging the superior inter-network collaboration capabilities of HWNs. These capabilities are crucial for capturing different patterns within or across edge networks, a critical requirement that previous architectures failed to meet. For instance, the work in~\cite{luo2020hfel} proposed the scheduling technique with two-level model aggregation for HFL in HWNs to reduce computation and communication costs. The authors in~\cite{xu2021adaptive} suggested an adaptive HFL system that integrated edge aggregation interval control and resource allocation to minimize the combined sum of training loss and latency. Liu et al. ~\cite{liu2022hierarchical} proposed a training algorithm for HFL that prioritizes communication efficiency using model quantization techniques. Despite their efforts to investigate HFL in HWNs, two significant challenges remain unaddressed. Firstly, the assumption that all data within these networks is labeled is violated since the unlabeled data represents a significant portion of the entire data for each worker and must be exploited to enhance model performance and expedite convergence. Secondly, there was an oversight of the unbalanced and non-IID data distribution during the training phase, resulting in a global model that could not capture data patterns effectively.

To address both statistical and resource allocation challenges, the studies in~\cite{sattler2020clustered,luo2021energy,gong2022adaptive,lu2023auction} explored the unbalanced and non-IID data challenge and aimed to minimize training time and energy consumption. For example, Sattler et al.~\cite{sattler2020clustered} proposed the CFL approach that tackles unbalanced and non-IID data problems by grouping workers based on the similarity of their data distributions into clusters, each with a specialized model. Duan et al.\cite{duan2021flexible} introduced a CFL technique that clusters workers by optimizing similarities and scales by adding devices during training. Wei et al.\cite{wei2023edge} developed the cFedFN framework to enhance visual classification in edge computing by minimizing weight divergence amid unbalanced and non-IID data. In~\cite{lu2023auction}, the authors explored an auction mechanism with mean-shift clustering for selecting workers based on data distribution, aiming to improve model convergence while reducing energy consumption. Within implementing CFL in HWNs, the study in~\cite{hamood2023clustered} proposed a clustered multitask federated distillation approach to address service heterogeneity, where local models have different structures. While these studies significantly contributed to advancing CFL approaches, they mainly focused on applying CFL with labeled data, representing only a limited portion of the entire data, using supervised learning in edge and hierarchical networks.

Integrating both SSL and FL to tackle the challenge of insufficient labeled data has garnered notable attention in recent research~\cite{albaseer2022semi,diao2022semifl,10159650}. For example,  Diao et al.~\cite{diao2022semifl} introduced a novel framework that combines the SSL with FL, enabling devices having only unlabeled data to participate in model training over multiple local epochs while the central server holds a limited set of labeled data. The work in~\cite{albaseer2022semi} developed the federated semi-supervised learning scheme, which considers various challenges in wireless communication, including bandwidth constraints, latency, and energy efficiency. In addition, Wang et al.~\cite{10159650} investigated personalized approaches in semi-supervised federated learning and used strategies such as adaptive worker variance reduction, application of local momentum, and normalized global aggregation to address device heterogeneity and facilitated the model convergence process.
 \begin{figure}[t]
\centering
  \includegraphics[width=1\linewidth]{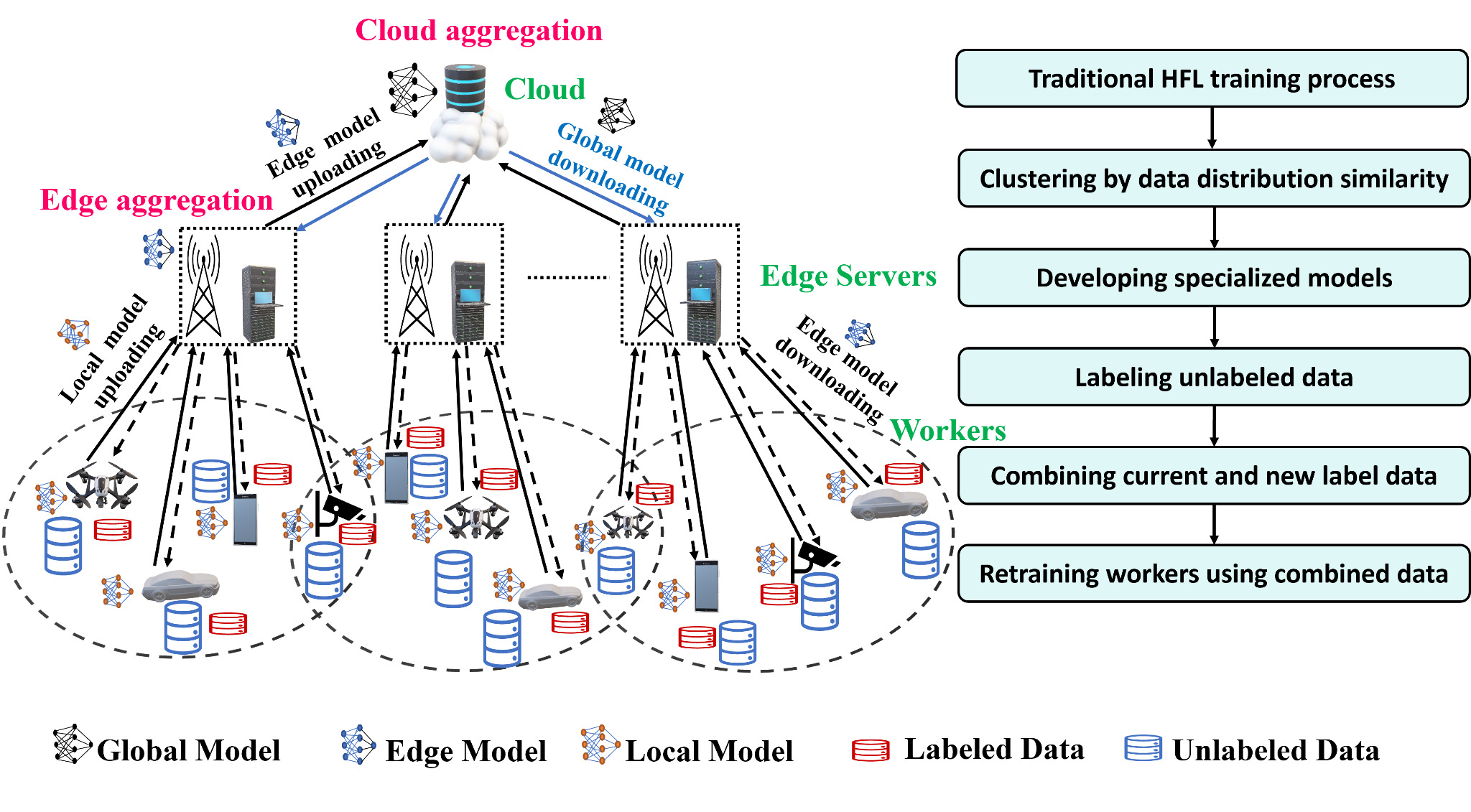}
\caption{The system model.}
\label{fig:figure1}
\end{figure}

Despite their contributions, critical gaps remain in CFL within HWNs, including the lack of combined CFL with SSL in environments dominated by unlabeled data. In addition, unbalanced and non-IID data challenges, device incompatibility, and limited network resources remain largely unresolved, especially with unlabeled data settings. These unresolved problems emphasize the need for a novel approach that leverages the vast unlabeled data alongside the small amount of labeled data while considering the system and data heterogeneities and resource consumption in HWNs. A more robust and reliable labeling and scheduling framework is essential to ensure the efficient operation of CFL systems with SSL in HWNs. 
\section{PRELIMINARIES}
\label{Sys_model}
In this section, we provide an overview of the system, HFL, and CFL models, explore the integration of CFL with SSL within HWNs, and analyze the processing and networking models.
\subsection{System Model}
As depicted in Fig. \ref{fig:figure1},  the system model includes three main layers: a cloud server at the top layer, a set of edge servers, $\mathcal{J}=\{1,\dots, J\}$ (where $J=|\mathcal{J}|$ representing the total number of edge servers), in the middle layer, and a set of workers,  $\mathcal{I}=\{1,\dots, I\}$ (with $I=|\mathcal{I}|$ indicating the total number of workers), at the bottom layer. Here, the cloud communicates with the edge servers, which connect to the assigned workers according to their physical proximity.
Each $i$-\textit{th} worker possesses a dataset, $\mathcal{D}_i$,  including a smaller subset of labeled data, $\mathcal{D}^l_i=\{{x}^{(i)}_{s,d}\in \mathbb{R}^d,~ y^{(i)}_{s}\}_{s=1}^{D^l_i}$, and a larger amount of unlabeled data, $\mathcal{D}^u_i=\{{x}^{(i)}_{s,d}\in \mathbb{R}^d\}_{s=1}^{D^u_i}$. Here, ${x}^{(i)}_{s,d}$ represents the feature vector with $d$ dimensions for the $s$-\textit{th} sample on the $i$-\textit{th} worker, and $y^{(i)}_{s}$ is the corresponding label. Hence, the entire dataset for each worker is a union of labeled and unlabeled data, $\mathcal{D}_i=\mathcal{D}^l_i\cup\mathcal{D}^u_i$, where ${D}^l_i=|\mathcal{D}^l_i|$ and ${D}^u_i=|\mathcal{D}^u_i|$ refer to the total number of labeled and unlabeled samples for the $i$-\textit{th} worker, respectively. Fig. \ref{fig:figure1} also illustrates the complete HFL process flow, beginning with traditional HFL training and progressing through the integration of labeled and unlabeled data, culminating in retraining workers using the combined dataset for enhanced model performance.
Table \ref{Tab:Notation} summarizes the main symbols used in this work.
\subsection{HFL Model}
\begin{table}[t]
\centering
\footnotesize
\caption{Main Symbols}
\label{Tab:Notation}
\begin{tabular}{p{1cm}p{6cm}} 
 \hline
$I$	& number of workers in the HWN \\ \hline
$J$	& number of edge servers in HWN \\\hline
$\mathcal{D}^{l}_i$ & the  labeled  data for $i$-\textit{th} worker\\\hline
$\mathcal{D}^{l}_i$ & the unlabeled data for $i$-\textit{th} worker\\\hline
${\mathcal{D}}$ & the total data samples for all workers in the HWN\\\hline
$\theta_j$ & the parameters of edge  model of $j$-\textit{th} edge server \\\hline
$\theta_r$ & the global model parameters of  the cloud at $r$-\textit{th} round \\\hline
$l_s(\theta_r)$ & the loss function for single data sample \\\hline
$F_i(\theta_r)$ & the local loss function for $i$-\textit{th} worker \\\hline
$\nabla F_i(\theta_r)$ & the local gradient for $i$-\textit{th} worker\\\hline
$N_i$ & number of local updates for $i$-\textit{th} worker \\\hline
$\eta$ & learning rate \\\hline
$\mathcal{P}$ & number of epochs \\\hline
$ sim^j_{i,i'}$& the cosine similarity between $i$ and $i'$ workers in the $j$-\textit{th} edge server \\\hline
$b$ & batch size \\\hline
$\mathcal{C}^j_r$ &  the available selected workers for edge server $j$\\\hline
$T^{cmp}_{i}$ & local computation time of $i$-\textit{th} worker\\\hline
$f_i$ & the local computation frequency at $i$-\textit{th} worker\\\hline
$\Psi$ & number of cycles required  to process one sample \\\hline
$\psi(i)$ & the data distribution of worker $i$\\\hline
$T_{budg}$ & the total time budget for the entire training process\\\hline
$T_r$ & deadline for the cloud in $r$-\textit{th} round\\\hline
$\beta_{j,i}$ & the bandwidth allocation ratio for $i$-\textit{th} worker\\\hline
$T^{com}_ {i}$ & the uploading time for updates from $i$-\textit{th} worker\\\hline
$r_i,r_j$ &  the achievable data rate required by the $i$-{th} worker and $j$-{th} edge server, respectively \\\hline
$\sigma$ & model size \\\hline
$\Phi$ & confidence threshold probability \\\hline
$\epsilon_1,\epsilon_2$ &  predefined thresholds to control clustering process\\
\hline
\end{tabular}
\end{table}
Designing HFL within HWNs is crucial to leveraging the diverse data patterns in edge networks and enabling collaborative knowledge exchange to develop a robust global model capable of capturing fine details under privacy constraints. The learning process of HFL is performed using FedAvg~\cite{mcmahan2017communication}, which can be detailed as follows: the cloud first plays a central role in optimizing the training process by managing model aggregation and resource allocation and maximizing the performance of the global model. During each training round, the cloud instructs edge servers to select a subset of available active workers, $\mathcal{C}^j_r$, for each $j$-\textit{th} edge server while gathering necessary information such as battery levels and processing capabilities. It also synchronizes updates with edge servers, adhering to deadline constraints to prevent delays.  The cloud then sends random global model parameters, $\theta_\circ$, to workers via edge servers, in which each worker initiates training using the locally labeled data and then uploads the updates to its edge server. Each edge server then aggregates and fuses these updates, developing a new version of an edge model. The cloud then aggregates and averages the edge models to update the global model, which is sent back to the workers for further training in the following rounds. This iterative process is repeated until the optimized global model converges to the optimal solution. It is worth noting that the local models are updated using stochastic gradient descent (SGD) with mini-batches by leveraging their labeled data. Updating these local models relies on several factors, including the volume of labeled data (${D}^l_i$), the time allotted for model uploads, and the computation frequency of the workers. Mathematically, the total number of updates executed during a single epoch for the $i$-\textit{th} worker's dataset using mini-batch SGD can be defined as follows:
\begin{equation}
\small
    N_i=\left\lceil\frac{{D}^l_i}{b}\right\rceil,
\end{equation}
where $b$ is the mini-batch size and $\left\lceil \cdot \right\rceil$ is a ceiling function that ensures the result is always an integer. To optimize the accuracy of local models throughout the training phase, each $i$-\textit{th} worker implements the mini-batch SGD algorithm $N_i$ times to reduce errors between actual and predicted labels at every $r$-\textit{th} round. Therefore, the local loss function for each $i$-\textit{th} worker can be given as:
\begin{equation}
\small
    F_i(\theta_r):=\frac{1}{D^l_i} \sum_{s \in \mathcal{D}^l_i} l_s(\theta_r),
\label{loss-func}
\end{equation}
where $l_s(\theta_r)$ is the local loss function for each $s$-\textit{th} sample.  Each $i$-\textit{th} worker seeks to minimize $l_s(\theta_r)$ within $\mathcal{D}^l_i$ using SGD, which iterates through each $s$-\textit{th} sample to reduce this loss function as follows:
 \begin{equation}
  l_s(\theta_r)= \mathcal{L}(x^{(i)}_{s,d},y^{(i)}_s; \theta_r). 
\end{equation}
This process results in an optimal local model that closely aligns with the distribution of its labeled data. Ultimately, the goal of HFL is to reduce the overall loss across all workers' labeled datasets ($\mathcal{D}^l = \bigcup_{i=1}^{I} \mathcal{D}^l_i$).
\subsection{Clustered Federated Learning (CFL)}
Without altering the communication protocols of traditional HFL, CFL can expand the HFL framework by tailoring it to specific subsets of workers with similar data distributions, effectively addressing the challenges of unbalanced and non-IID data. More specifically, CFL splits workers into clusters, each with a specialized model that fits the cluster's data distribution well, ensuring a more efficient learning process.

\noindent\textit{\textbf{Assumption 1 (CFL):}  For each $j$-th edge server within HWNs, a subset of workers $\mathcal{I}_j$, derived  from $K$ different data generation distributions: \{$\psi_1,\dots, \psi_K$\}, with ($K \leq I_j$), is classified into $M_j$ clusters: $\mathcal{M}_j=\{g_{j1},\dots, g_{jm}\}$, where $\bigcup_{n=1}^{M_J} g_{jn} = \{1, ..., I_j\}$, in such a way that each worker subset $g \in \mathcal{M}_J$ aligns with the traditional HFL criteria.}

Here, $M_j=|\mathcal{M}_j|$ refers to the number of clusters within the $j$-\textit{th} edge network; thus, the total number of clusters across all edge networks in HWNs is given as $M=\sum_{j=1}^{J} M_J$. The CFL process can be visualized as a parameterized tree structure, where HFL forms the root and clusters of workers represent this tree's branches. The learning process of CFL is initiated by training and uploading updates to edge servers. Subsequently, edge servers compute the similarity between gradient updates for each pair of workers, determining the necessity for their partitioning into different clusters. This process continues until the training in each edge network reaches the stationary point ($\theta^*_j$), where no further enhancements in the edge model's performance are feasible with the existing dataset. As a result, organizing workers into clusters is crucial to improve model performance and accuracy by providing each cluster with a specialized model perfectly adapted to its specific dataset. In CFL, the cosine similarity ($sim$) is used to verify the gradient similarity between any two workers within the same $j$-\textit{th} edge network or across different networks. Formally, the cosine similarity is defined as follows:

\begin{footnotesize}
\begin{align}
sim^j_{i,i'}&:=sim(\nabla F_i(\theta^*_j), \nabla F_{i'}(\theta^*_j)):=\frac{\langle  \nabla F_i(\theta^*_j),\nabla F_{i'}(\theta^*_j)\rangle}{\parallel\nabla F_i(\theta^*_j) \parallel \parallel\nabla F_{i'}(\theta^*_j)\parallel} 
\nonumber \\ &= 
\begin{cases}
1, & \text{if $\psi(i)=\psi(i')$}\\
-1, & \text{ if $\psi(i)\neq \psi(i')$},
\end{cases}
    \label{similarityformula}
\end{align} 
\end{footnotesize} 
Here, $\psi(i)$ and $\psi(i')$ are the data-generating distribution of $i$ and $i'$ workers, respectively. Note that $sim$ takes the value of either $1$ or $-1$ in (\ref{similarityformula})  for two reasons: first, it is computed in (\ref{similarityformula}) in the extreme point of the HFL objective (i.e., stationary point,$\theta^*_j$) where gradient vectors for workers are either in completely opposing ($-1$) or aligned directions ($1$); second,  this binary separation simplifies the decision-making process by distinguishing between aligned and opposed gradient vectors. Therefore, the correct bipartitioning is obtained as $g_{j1}=\{i|sim^j_{i,0}=1\}$ and  $g_{j2}=\{i|sim^j_{n,0}=-1\}$. To achieve this, the correct partition for $\mathcal{I}_j$ in each $j$-\textit{th} edge server into different clusters must meet the following conditions: First, the solution of HFL objective gets closer to the stationary point as follows:
\begin{equation}
\small
  0  \leq \biggl\|\sum_{i\in \mathcal{I}_j}\frac{D^l_i}{{D}^l} \nabla_{\theta_j} F_i(\theta^*_j)\biggr\| < \varepsilon_1.\label{cond1}
\end{equation}
Second, the distance between individual workers and the stationary point of their local loss function must meet the following:
\begin{equation}
\max_{i \in \mathcal{I}_j} \left\| \nabla_{\theta_j} F_i(\theta^*_j)\right\|>\varepsilon_2>0,
\label{cond2}   
\end{equation}
where $\varepsilon_1$ and $\varepsilon_2$ are predefined hyperparameters used to control the clustering process within unbalanced and non-IID data settings, significantly impacting the convergence and performance of the proposed approach. Specifically, $\varepsilon_1$ evaluates clustering quality as learning progresses towards the stationary point, while $\varepsilon_2$ controls the grouping of workers based on client availability and data heterogeneity. For instance, setting $\varepsilon_1$ too high may result in early clustering, combining different data distributions and reducing model performance. Similarly,  if $\varepsilon_2$ is set too large, it may result in overly broad clusters, compromising the model's ability to learn from detailed variations within the data. It is worth noting that once conditions (\ref{cond1}) and (\ref{cond2}) are satisfied, edge servers upload their specialized models to the cloud for further processing. The cloud then computes the cosine similarity between specialized models from different edge networks to enhance generalization and robustness. Model training within each cluster continues using the labeled data until the HFL objective either encounters a new stationary point that necessitates a split or reaches a stopping point where workers have congruent data distributions (i.e., no further split is required).
\subsection{Communication and Computation Models}
\noindent\textbf{Computation model:}
During this phase, each edge server sends its model parameters to the assigned workers to initiate the training, requiring  time and energy. Assuming that $\Psi$ represents the number of CPU cycles required to process a single data sample and $f_i$ is the computation frequency, the computation time for the $i$-\textit{th} worker to execute the local solver across $\mathcal{P}$ epochs during a given round, $r$, can be calculated as follows:
\begin{equation}
\small
    T^{cmp}_i=\mathcal{P}\frac{{D}^l_i\Psi}{f_i}.
    \label{local_comp_t}
\end{equation}
 Hence, the computation energy for the $i$-\textit{th} worker can be defined as:
\begin{equation}
\small
 e_{i,j}^{cmp}=\frac{\Upsilon_i}{2} (\mathcal{P} f_i^2  {D^l_i \Psi}),
   \label{energy_cmp_n_1}
\end{equation}
where $\Upsilon_i$ is the capacitance coefficient for the energy efficiency of the $i$-\textit{th} worker's processor.\\
\textbf{Communication model:} Participating workers in this phase upload their models to their edge servers via wireless channels. Given the available spectrum constraints, we use orthogonal frequency-division multiple access (OFDMA) to optimize communication within HWNs. This technique divides the spectrum into multiple orthogonal sub-channels, each uniquely assigned to an $i$-\textit{th} worker, enabling simultaneous data transmissions for workers across HWNs. Supposing the channel gain between the $i$-\textit{th} worker and $j$-\textit{th} edge server refers to ${h}_{i,j}$. Thus, the transmission data rate for $i$-\textit{th} worker to send its local models is defined as:
\begin{equation}
\small
  {\delta^{com}_i = \beta_{i,j}~B~\text {log}_2\left(1 + \frac{| {h}_{i,j}|^2 P_i}{ N_\circ }\right)}, 
    \label{Data_rate}
\end{equation}
where $\beta_{i,j}$ represents the portion of bandwidth allocated to the $i$-\textit{th} worker out of the total system bandwidth ($B$), $N_\circ$ is the total noise power of complex additive white Gaussian noise (AWGN) across the entire bandwidth, and $P_i$ indicates the transmission power of the $i$-\textit{th} worker. Hence, the time and energy required for the worker $i$ to transmit its local model of size $\sigma$ to the edge server $j$ are computed, respectively, as:
\begin{equation}
 T^{com}_i=\frac{\sigma}{\delta^{com}_i},  
\end{equation}
\begin{equation}
    e_{i,j}^{com}=T^{com}_i P_i.
    \label{comm_energy_n}
\end{equation}
Given the various challenges participant workers face, including handling extensive datasets or operating with limited resources, the time it takes to complete their tasks may exceed that of their peers. These workers (i.e., stragglers) can significantly delay the training process. To mitigate this issue, each edge server, coordinating with the cloud, sets a deadline ($T^{i}_{C^j_r}$) for its assigned workers at every $r$-\textit{th} round to prevent long waiting times.
 Thus, the total time required for the $i$-\textit{th} worker to train and upload models to the $j$-\textit{th} edge server is defined as:
\begin{equation}
\small
    T_{\mathcal{C}^j_r}^{i}=\max \left( T^{cmp}_i+ T^{com}_i\right).
\end{equation}
Thus, the total energy required for a set of workers, $\mathcal{C}_r^j$, during the $r$-\textit{th} round can be expressed as:
\begin{equation}
\small
E_{\mathcal{C}_r^j}^{i}=\sum^{\mathcal{C}_r^j}_{i=1}e_{i,j}^{cmp}+e_{i,j}^{com}.
    \label{Total_energy_edge}
\end{equation}
Similarly, the time required for the $j$-\textit{th} edge server to upload its edge model to the cloud is given as: 
\begin{equation}
\small
    T^{\mathrm{cld}}_{j} = \frac{\sigma}{\delta^{com}_{j}},
\end{equation}
where $\delta^{com}_j$ is the data rate at $j$-\textit{th} edge server for uploading edge and specialized models to the cloud.

Ultimately, the overall total time and total energy required by workers to complete one global round are respectively calculated as follows:
\begin{equation}
\small
    T_r=\underset{j\in \mathcal{J}} \max ~(T^{\mathrm{cld}}_{j}+T_{\mathcal{C}^j_r}^{i})
\end{equation}
\begin{equation}
\small
   E_r=\sum_{j=1}^{J} (E_{\mathcal{C}_r^j}^{i}+E^{\mathrm{cld}}_{j}),
\end{equation}
respectively, where $E^{\mathrm{cld}}_j$ is the total energy required for $j$-\textit{th} edge server to transmit its models to the cloud.
\section{Problem Formulation}
\label{problem_form}
As stated before, a large portion of the data for each $i$-\textit{th} worker is unlabeled, necessitating the need to use a pseudo-labeling technique to improve model performance. To this end, the proposed approach equips each cluster with a specialized model tailored to its specific labeled data distribution. However, a critical challenge arises when these specialized models label the unlabeled data from different distributions, leading to inaccuracies in labeling. This dilemma emphasizes the need for alternative strategies to refine the labeling process effectively. In addition, the optimal timing for using these models complicates the process further. Moreover, wireless network constraints, such as non-IID data, worker heterogeneity, and limited bandwidth, restrict full worker participation in each training round.  In response, we aim to efficiently leverage the abundance of unlabeled samples to enhance overall system performance.

For an effective labeling process, we set a confidence threshold, $\Phi$, to select labels only when predictions exceed the threshold's probability, ensuring the reliability of selected labels. In other words, $\Phi$ is a key variable in the training process, determining which pseudo-labels are accepted or rejected based on application needs. For example,  a higher $\Phi$ is vital for applications such as medical image analysis, ensuring only highly confident predictions are used to enhance reliability. Conversely, a lower $\Phi$ broadens decision-making by accepting more predictions but risks, including less accurate ones, leading to degradation in the performance.
 Mathematically, to achieve flexible labeling using $\Phi$, the following condition must be met:
\begin{equation}
\small
    \max(\mathrm{Prob}(\hat{y}_a)) \geq \Phi.
\end{equation}
Here, $\mathrm{Prob}(\cdot)$ represents the class output probability, and $\hat{y}_a$ is the pseudo-label predicted for the $a$-\textit{th} sample by prediction models. Thus, a label with the highest probability (i.e., the top-class label) for the $a$-\textit{th} sample is defined as follows:
\begin{align}
\small
\hat{y}^{*}_a=\max(\mathrm{Prob}(\hat{y}_a)).
\end{align}
The effect of $\Phi$ on the loss function $F(\theta_i)$ for the $i$-\textit{th} workers is measured by combining the loss from the labeled data $F_l(\theta)$ and the loss from the pseudo-labeled data $F_{psdo}$ as follows:

\begin{footnotesize}
    \begin{align}
    F(\theta_i,\mathcal{D}^l_i,\mathcal{D}^l_{psdo,i})&=\frac{1}{D^l}\sum^{D^l}_{s=1}\mathcal{L}(x^{(i)}_{s,d},y^{(i)}_s)+\nonumber\\ 
    &\frac{1}{{D}^l_{psdo,i}}\sum^{D^l_{psdo,i}}_{a=1}\mathcal{L}(x^{(i)}_{a,d},\hat{y}^{(i)}_a;\theta),
\end{align}
\end{footnotesize}
where ${D}^l_{psdo,i}$ is the number of pseudo-labeled samples meeting the confidence threshold. Accordingly, the total number of labeled data samples after combining pseudo-labels and the training samples is defined as:
\begin{equation}
\small
D^{l}_{i,new}= D^l_i+{D}^l_{psdo,i}.
\label{new_dataset}
\end{equation}
From (\ref{new_dataset}), the edge networks incur more training time due to the increase in labeled data for each $i$-\textit{th} worker, which requires adapting its computation time to satisfy the deadline constraints. Mathematically, the new computation time for the $i$-\textit{th} worker can be obtained as:
\begin{equation}
\small
    T^{cmp}_{i,new}=P\frac{{D}^{l}_{new,i}\Psi}{f_i}.
    \label{new_time}
\end{equation}
 As a result, we aim to develop an optimal framework that leverages specialized models to effectively label the unlabeled data in non-IID environments, considering resource constraints. Our main goal is to minimize the loss function  $F(\theta_i,\mathcal{D}^l_i,\mathcal{D}^u_i)$  for each worker using labeled and pseudo-labeled data to achieve better model performance and efficient resource consumption. This can be achieved by designing optimal prediction models and time schemes to exploit the unlabeled data effectively and developing a scheduling scheme that selects workers based on better resources or equitable distribution of resources. Let $R$ be the total number of global rounds constrained by the time budget $T_{bdgt}$, $\mathcal{C}_{[R]}=[\mathcal{C}_{1},\dots, \mathcal{C}_{R}]$  be the selected scheduling sets for all rounds, and $T_{[R]}=[T_1,\dots, T_R]$ be the corresponding deadlines for each round. Thus, the optimization problem can be formulated as follows:

\begin{subequations}
\footnotesize
\begin{align}
\textbf{P1:} \underset{
\substack{\boldsymbol{\theta}, R,T_{[R]}, \mathcal{C}_{[R]}}
}{\text{min}}
&~\sum_{r=1}^{R} \sum_{i=1}^{I} F(\theta_i,\mathcal{D}^l_i,\mathcal{D}^u_i) 
    \label{eq:OptmizedProblem1}\\
\textrm{s.t.} \quad\nonumber \\ 
& F(\theta_{m}) - F(\theta^*_{m})\leq\epsilon, \quad \forall m \in \mathcal{M}, \label{eq:convergence_constraint} \\
& \sum_{r}^{R} T_r(\mathcal{C}_{[R]})\leq T_{bdgt}, \label{total_time}\\
&(T^{cmp}_{i,new}+T^{com}_{i}) \leq  T_{\mathcal{C}^j_r}^{i},~\forall i \in \mathcal{I}, j \in \mathcal{J},
\label{Deadline_one_client}\\
&  P_i^{min}  \leq  P_i \leq  P_i^{max}, \quad \forall i \in \mathcal{I},\label{Power_const}\\
&  f_i^{min} \leq   f_i \leq  f_i^{max}, \quad \forall i \in \mathcal{I},\label{freq_const}\\
& \sum_{i=1}^{I} B(\theta_i) \leq B, \label{bw_constraint}\\
& A^j_i \in \{0,1\},\quad \forall i \in \mathcal{I},~\forall j \in \mathcal{J}.
\label{association}
\end{align}
\end{subequations}
Constraint (\ref{eq:convergence_constraint}) guarantees that each $m$-\textit{th} specialized model converges to the optimal solution using the labeled and pseudo-labeled data. Constraint (\ref{total_time}) ensures that the entire training process for all selected workers in all edge networks within HWNs is finished within the predefined time allocation, $T_{bdgt}$. Constraint (\ref{Deadline_one_client}) ensures that the maximum training and uploading time in round $r$ does not exceed the deadline $T_{\mathcal{C}^j_r}^{i}$ set by each $j$-\textit{th} edge server. Constraint~(\ref{Power_const}) bounds the transmission power for each $i$-\textit{th} worker within a minimum and maximum limit to optimize energy consumption. Constraint~(\ref{freq_const}) limits the CPU speed for the $i$-\textit{th} worker to balance computational efficiency and energy usage. Constraint~(\ref{bw_constraint}) ensures that the total bandwidth used by all models across all workers does not exceed the system's available bandwidth ($B$). Constraint~(\ref{association})is a binary constraint that indicates whether $i$-\textit{th} worker is associated with an $j$-\textit{th} edge server, ensuring proper communication flow between workers and edge servers (i.e., $1$ if the worker connects to a specific edge server, $0$ if not).

For the following reasons, \textbf{P1} is an intractable optimization problem, characterized by its mixed-integer nonlinear programming (MINLP) structure and NP-hard complexity.  This arises from the need to determine how each parameter, including $R$, $\boldsymbol{\theta}$, and $\mathcal{C}_{[R]}$  influences the weight vector associated with each worker's model, $ F(\theta_i,\mathcal{D}^l_i,\mathcal{D}^u_i)$. In addition, labeling unlabeled data requires careful model and time selections.  The complexity of the problem is further increased not only because of the unexpected changes in wireless channel conditions and the different computation times among workers during the training phase but also because of the difficulties of HWNs, such as the two levels of model aggregation and computations. All these challenges complicate selecting the most effective model scheme, determining the optimal timing, and choosing the best worker, thus making it difficult to achieve a straightforward solution. 
\section{Proposed Solution}
\label{proposed_sol}
\begin{figure}[t]
    \centering
    \includegraphics[width=1\linewidth]{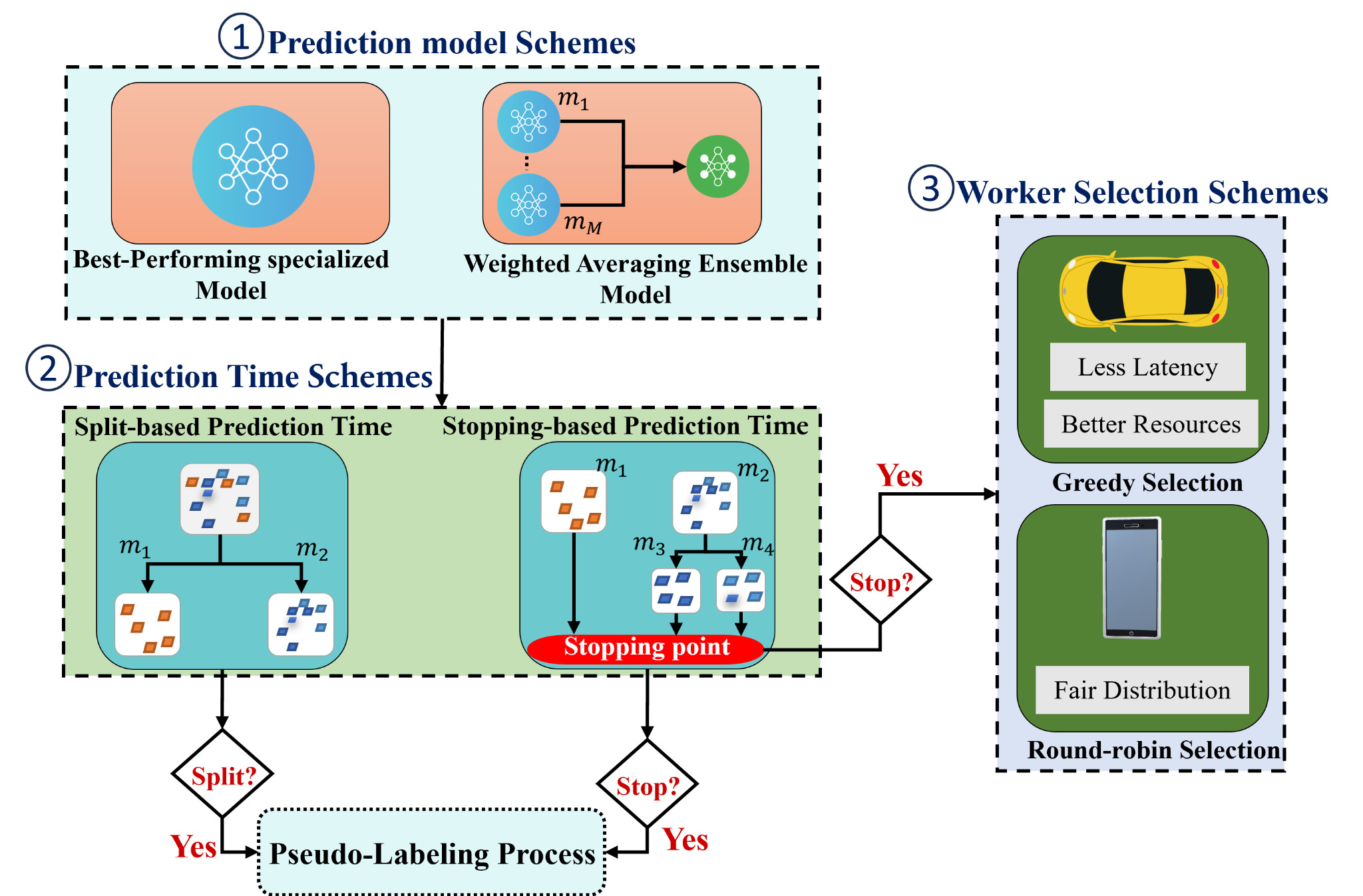}
    \caption{System framework diagram for the CFSL approach.} 
    \label{CFSL_Flowchart}
\end{figure}

In this section, we introduce our tractable solution, including the prediction model and time schemes as well as the worker selection strategy to ensure proper labeling while considering network limitations. However, a critical research challenge arises in optimizing specialized models to accurately label data and determining the optimal timing for deploying these schemes to maximize accuracy and resource efficiency in complex and limited-resource HWNs. To address this, our proposed approach effectively handles the labeling process under non-IID data settings and optimizes the resource consumption of computation and communication models. It employs two main schemes: the best-performing specialized model and the weighted-averaging ensemble model for labeling unlabeled data. Further, our approach uses split-based and stopping-based prediction times to determine the optimal label initiation timing.  Lastly,  network efficiency is also optimized through two scheduling schemes, greedy and round-robin selections, to reduce time delay and energy consumption, resulting in cost-efficient HWN systems. Fig. \ref{CFSL_Flowchart} illustrates the system framework, clearly defining the network structure and the novel schemes implemented in HWN environments.

Since \textbf{P1} is intractable,  using traditional optimization methods to solve this problem is impractical. Thus, the heuristic-based solution introduces two prediction models, prediction times and worker selections, to optimize model performance and network parameters. \textbf{P1}  can be reformulated into \textbf{P2} and \textbf{P3}, each focusing on a specific model prediction as detailed in \ref{Predi_models}. These sub-problems aim to minimize the loss function $F(\theta_i,\mathcal{D}^l_i,\mathcal{D}^u_i)$ while maximizing the utility $\mathcal{U}(\theta_m, \mathcal{D}^u_i)$ in labeling.  We extensively detail the proposed approach's schemes as follows:
\subsection{Prediction model schemes}
\label{Predi_models}
\begin{figure}[t]
\centering
  \includegraphics[width=1\linewidth]{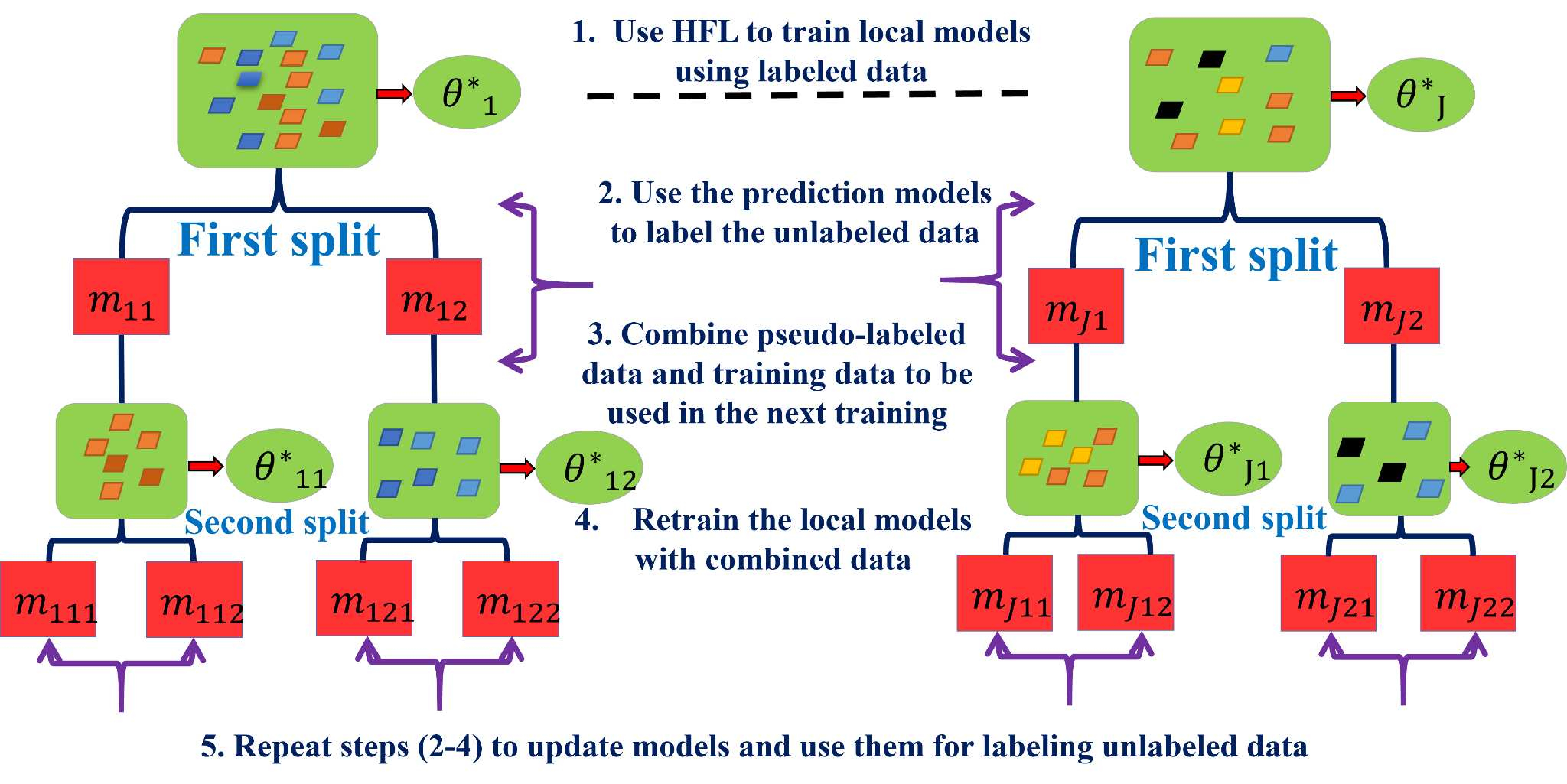}
\caption{Illustrating labeling process for the CFSL in HWNs.}
\label{fig:figure2}
\end{figure}

As previously noted,  we propose prediction model schemes, including the best-performing specialized and weighted-averaging ensemble models for predicting labels. Fig. \ref{fig:figure2} illustrates the labeling process in our proposed approach. This figure shows that both prediction model schemes use the pseudo-labeling technique to label the unlabeled data by selecting only labels that pass $\Phi$, ensuring the most reliable label selection. 

Comparing our proposed prediction models with scenarios that have only a single neural network is crucial. The proposed approach outperforms the single neural networks for the following reasons: First, the single neural network has a limited ability to capture nuanced data details in non-IID settings, while our specialized models are tailored to different data distributions, improving accuracy in labeling within each cluster. Second, a single model applies uniform labeling across all workers, leading to suboptimal results, while our adaptive schemes select the most effective model for each worker, improving label accuracy. Finally, the single neural network in non-IID settings struggles with conflicting data patterns, whereas our multiple specialized models are better suited to handle these distributions, ensuring more accurate labeling.
\subsubsection{Best-performing specialized model}
In this scheme, each $i$-\textit{th} worker select the specialized model with higher labeling accuracy and minimum labeling latency, providing better quality of labeling. In CFSL, the critical challenge lies not only in accurately labeling unlabeled data from the same distribution but also in effectively handling data originating from different distributions. To evaluate this scheme's effectiveness on the unlabeled, unseen data, we aim to select the best-performing specialized model that minimizes the worker's loss function $F(\theta_i,\mathcal{D}^l_i,\mathcal{D}^u_i)$ on labeled data while maximizing its utility function, $\mathcal{U}(\theta_m, \mathcal{D}^u_i)$, in labeling the unlabeled data during each $r$-\textit{th} round. Thus, \textbf{P1} can be reformulated as follows:

\begin{subequations}
\footnotesize
\begin{align}
\textbf{P2:}\underset{
\substack{\boldsymbol{\theta},R,\mathrm{z},T_{[R]}, \mathcal{C}_{[R]}}}{\text{min}}
&~\sum_{r=1}^{R}
    \sum_{i=1}^{I} \left[ F(\theta_i,\mathcal{D}^l_i,\mathcal{D}^u_i) -
    \lambda  \sum_{m \in \mathcal{M}}  z_{i,m}\mathcal{U}(\theta_m, \mathcal{D}^u_i)\right] \label{eq:OptmizedProblem1}\\
\textrm{s.t.} \quad\nonumber \\ 
&\mathrm{const.~ (\ref{eq:convergence_constraint})-(\ref{association})} \label{collaction1}\\
& \sum^{M}_{m=1} z_{i,m}=1,\quad \forall i \in \mathcal{I}, \label{binary22}\\
& z_{i,m} \in \{0,1\},\quad \forall i \in \mathcal{I},~\forall m \in \mathcal{M},
\label{binary222}
\end{align}
\end{subequations}
where $\lambda$ is a trade-off parameter. Constraint~(\ref{binary22}) stipulates to ensure that each $i$-\textit{th} worker uses only one specialized model $\theta_m$ out of $M$ specialized models for the pseudo-labeling process. Constraint~(\ref{binary222}) is a binary variable denoting whether the $m$-\textit{th} specialized model is used to label the unlabeled data of $i$-\textit{th} worker ( $z_{i,m}=1$ if yes, $0$ if not). 

In this scheme, the best-performing specialized model is selected as follows: After receiving $\theta_\circ$ from the cloud, workers train local models using labeled data and upload updates to their edge servers. The edge servers compute cosine similarity among these updates, cluster workers based on conditions~(\ref{cond1}) and (\ref{cond2}), and then send the resulting specialized models to the cloud. The cloud performs a further similarity computation among the specialized models to capture different patterns across edge networks. As a result, $M$ specialized models are developed, where each $i$-\textit{th} worker selects the best-performing specialized model for labeling the unlabeled data and injects pseudo-label samples into the existing training data, forming a combined dataset used for the next training rounds.

Note that the best-performing specialized model for the $i$-\textit{th} worker might not be the one was initially designed for it during the training; instead, it could be the model developed in a different worker within the same network or even from different edge networks. This stems from the fact that this specialized model may fail to accurately label the unlabeled data for its workers under non-IID data settings. Therefore, the $i$-\textit{th} worker communicates with the cloud, requesting access to all specialized models developed within HWNs. This allows each $i$-\textit{th} worker to evaluate the utility of each model for accurate labeling, selecting the model that not only offers the highest labeling accuracy but also minimizes labeling latency, ensuring an efficient and accurate labeling process.
\subsubsection{ Weighted-averaging ensemble model}
In this scheme, we aim to exploit the capabilities of all specialized models by combining them at every round to develop a unified model used in the labeling by all workers. Here, $\alpha_{i,m}$ plays a critical role in this scheme by assigning specific weight to each $m$-\textit{th} specialized model. Precisely, the specialized models that contribute more to the unified model attain more weights than those with low performance, ensuring fairness and enhancing the overall prediction accuracy. This weight is adjusted heuristically to improve label accuracy by prioritizing the most reliable models in the ensemble.  Thus, we reformulate \textbf{P1} for this scheme as follows:

\begin{subequations}
\footnotesize
\begin{align}
\textbf{P3:}\underset{
\substack{\boldsymbol{\theta},R,\boldsymbol{\alpha},T_{[R]}, \mathcal{C}_{[R]}}}{\text{min}}
&~\sum_{r=1}^{R}
    \sum_{i=1}^{I} \left[ F(\theta_i,\mathcal{D}^l_i,\mathcal{D}^u_i ) -
    \lambda  \sum_{m \in \mathcal{M}}  \alpha_{i,m}\mathcal{U}(\theta_m, \mathcal{D}^u_i)\right] \label{eq:OptmizedProblem1}\\
\textrm{s.t.} \quad\nonumber \\ 
&\mathrm{const.~ (\ref{eq:convergence_constraint})-(\ref{association})} \label{collaction1} \\
& \sum^{M}_{m=1} \alpha_{i,m}=1,\quad \forall i \in \mathcal{I}, \label{binary2}\\
&   0\leq \alpha_{i,m}\leq 1,\quad \forall i \in \mathcal{I},~\forall m \in \mathcal{M},
\label{binary3}
\end{align}
\end{subequations}
where constraints~(\ref{binary2}) and~(\ref{binary3}) represent normalized weighted-averaging of all the specialized models.

Once the specialized models are uploaded to the cloud, it provides workers access to all specialized models. Each model generates pseudo-labels for the unlabeled data on each $i$-\textit{th} worker. These pseudo-labels are then combined using a weighted average, where the weights are determined by the models' utility and accuracy in labeling. The final prediction for each data sample is a weighted combination of the predictions from the different models, with higher utility models having more influence on the final label. This process repeats at each $r$-\textit{th} round to refine and enhance accuracy, ensuring that pseudo-labeling remains adaptive and effective.
\subsection{Prediction time schemes}
\begin{algorithm}[t!]
\footnotesize
\DontPrintSemicolon
\caption{Split-based Prediction Time Scheme}
\label{CFL1}
\KwIn{Device count $I$, initial parameters $\theta_0$, confidence threshold $\Phi$, and controlling parameters ($\varepsilon_1$ and $\varepsilon_2$)}
\SetKwFunction{CFLSteps}{CFLSteps}
\SetKwProg{Fn}{Function}{:}{}
\Fn{\CFLSteps{$\mathcal{M}$, $\varepsilon_1$, $\varepsilon_2$}}{
    \For{$g \in \mathcal{M}$}{
        Edge server gets $\Delta\theta_m \leftarrow \frac{1}{|g|} \sum_{i\in g} \Delta \theta_i$\\
        \If{$||\Delta \theta_g|| < \varepsilon_1$ \textup{and} $\max_{i \in g }||\Delta \theta_i|| > \varepsilon_2$}{
            $sim^j_{i,i'} \leftarrow \frac{\langle \Delta\theta_i, \Delta\theta_{i'}\rangle}{||\Delta\theta_i||||\Delta\theta_{i'}||}$\\
            $g_1,g_2 \leftarrow \arg\min_{g_1\cup g_2=g} \left(\max_{i\in g_1, i'\in g_2} sim^j_{i,i'}\right)$\\
            $\gamma_i \leftarrow \frac{||\nabla F_{\psi(i)}(\theta_j^*)-\nabla F_i(\theta_j^*)||}{||\nabla F_{\psi(i)}(\theta_j^*)||}$\\
            \If{$\max(\gamma_i) < \sqrt{\frac{1-sim^{\max}_{\mathrm{cross}}}{2}}$}{
                $\mathcal{M}_{\mathrm{tmp}} \leftarrow (\mathcal{M}_{\mathrm{tmp}}\setminus g) \cup \{g_1, g_2\}$\\
            }
        }
    }
}
\tcp{Checking the splitting conditions \ref{cond1} and \ref{cond2}}
\If{$M \geq 1 $}{
    Labeling $\mathcal{D}^u_i$ using the best-performing specialized model, $m$, or the weighted-averaging ensemble model\\
    \tcp{Meeting confidence threshold}
    \If{$\max(\mathrm{Prob}(\hat{y}_a)) \geq \Phi$}{ Injecting pseudo-labeled samples into the training data}
}
\tcp{Performing post-processing clustering}
\CFLSteps{$\mathcal{M}$, $\varepsilon_1$, $\varepsilon_2$}\par
$\mathcal{M} \leftarrow \mathcal{M}_{\mathrm{tmp}}$
\end{algorithm}
We provide  the split-based and stopping-based prediction time schemes, improving labeling efficiency while minimizing training time and energy consumption.
\subsubsection{Split-based prediction time}
This scheme initiates label prediction for unlabeled data once any edge server performs at least one split. In other words, once the specialized models are developed in our proposed approach, we label the unlabeled data using the two prediction model schemes. Algorithm~\ref{CFL1} describes the proposed approach mechanism using split-based prediction time.
\subsubsection{Stopping-based prediction time}
In this scheme, we begin the label prediction once clusters reach their stopping points.  We allow specialized models to perform further training until each cluster reaches its stopping point.  Mathematically, the stopping point can be defined as follows:
\begin{equation}
\small
\max_{i \in g_j} \left\| \nabla_{\theta_j} F_i(\theta^*_j)\right\|<\varepsilon_2.
\label{stopping}
\end{equation}
Once clusters meet (\ref{stopping}), workers directly predict labels and inject the pseudo-labeled samples into the training data to perform further training. This improves label prediction since the proposed approach waits for the specialized model stability before starting label prediction, reducing the risk of early or inaccurate labeling. Algorithm~(\ref{CFL2}) presents the CFSL with the stopping-based prediction time scheme.

The trade-off between the two prediction time schemes is more evident, emphasizing the balance between accuracy and resource efficiency. The stopping-based scheme prioritizes accuracy by aligning labeling with stable models, while the split-based scheme initiates labeling earlier during worker splitting to optimize resource consumption. This ensures that the labeling process remains effective and balanced, depending on the specific priorities of the system.
\subsection{Worker Selection Schemes}
The proposed approach utilizes two effective strategies, greedy and round-robin selection, to select workers in clusters that reach the stopping point. 
\subsubsection{Greedy selection scheme}
\begin{algorithm}[t!]
\footnotesize
\DontPrintSemicolon
\caption{Stopping-based Prediction Time Scheme}
\label{CFL2}
\KwIn{Device count $I$, initial parameters $\theta_\circ$, confidence threshold $\Phi$, controlling parameters $\varepsilon_1$ and $\varepsilon_2$}

\SetKwFunction{CFLAteps}{CFLSteps}

\tcp{Checking condition (\ref{stopping})}
\If{$\max_{i \in g_j} \left\| \nabla_{\theta_j} F_i(\theta^*_j)\right\| < \varepsilon_2$}{
    Labeling $\mathcal{D}^u_i$ using the best-performing specialized model ($m$) or the weighted-averaging ensemble model\\
    \tcp{Satisfying confidence threshold}
    \If{$\max(\mathrm{Prob}(\hat{y}_a)) \geq \Phi$}{ Injecting pseudo-labeled data into the training data}
}
\tcp{Developing specialized models}
\CFLSteps{$\mathcal{M}$, $\varepsilon_1$, $\varepsilon_2$}

Update $\mathcal{M}$
\end{algorithm}
Whenever clusters reach the stopping point, the greedy selection technique selects the best worker with less latency and better resources to complete the training until the model converges. This scheme significantly reduces training time and resource consumption, as we only care about the best workers for further training.
\subsubsection{Round-robin selection scheme}
In this scheme, once clusters approach stopping points, we schedule workers based on the fairness of the resources. Each worker in a specific cluster obtains the same opportunity to participate in the training regarding time and resources. Thus, we ensure equal contribution for all participant workers during training, which facilitates fair resource allocation and enables accurate label prediction across different data sources.
\subsection{ Integrated schemes}
\subsubsection{Best-performing specialized model with split-based prediction time and greedy selection (BMSPGS)}
\label{Scenario1}
As the name implies, this combined scheme uses the best-performing specialized model for labeling and initiates this scheme when edge servers perform at least one split. We perform training and labeling even after every cluster reaches its stopping point. In this phase, we use the greedy selection algorithm to select workers with less latency and better resources for further training and labeling. This process continues until the specialized models converge and unlabeled samples are accurately labeled.
\subsubsection{Best-performing specialized model with split-based prediction time and round-robin selection (BMSPRR)}
\label{Scenario2}
In this scheme, the method aligns with the scheme~(\ref{Scenario1}) in prediction model and timing schemes, where the only difference lies in the worker selection algorithm. This scheme utilizes the round-robin algorithm, cyclically allocating resources to ensure workers' equitable distribution.
\subsubsection{Best-performing specialized model with stopping-based prediction time and greedy selection (BMSTGS)}
\label{Scenario3}
The training process in this scheme is similar to the previous two schemes regarding the prediction model and worker selection schemes. The only variation is the time initiated for label prediction and pseudo-label sample injection. We delay the labeling process until each cluster reaches its stopping point. Accordingly, we label the unlabeled data and inject pseudo-samples into the training data for further fine-tuning.
\subsubsection{Best-performing specialized model with stopping-based prediction time and round-robin selection (BMSTRR)}
\label{Scenario4}
The main difference between this scheme and that in  \ref{Scenario3} lies in the worker selection mechanism. Here, we utilize the round-robin algorithm, ensuring a fair distribution of resources.
\subsubsection{Weighted-averaging ensemble model with split-based prediction time and greedy selection (EMSPGS)}
\label{Scenario5}
In this scheme, the weighted-averaging ensemble model labels unlabeled data and injects confident labels into the training data whenever at least one split is executed on edge networks.  Also, worker selection adheres to a greedy selection scheme, prioritizing workers with low latency and superior resources for participation.
\subsubsection{Weighted-averaging ensemble model with split-based prediction time and round-robin selection (EMSPRR)}
\label{Scenario6}
The training and labeling protocols in this scheme closely resemble those described in~\ref{Scenario5}, where the exception is the strategy of worker selection. Specifically, this scheme employs the round-robin algorithm for selecting workers.
\subsubsection{Weighted-averaging ensemble model  with stopping-based prediction time and greedy selection (EMSTGS)}
\label{Scenario7}
\begin{algorithm}[t!]
\footnotesize
\caption{The proposed Framework}
\label{CFL3}
\KwIn{Device count ($I$), initial parameters ($\theta_\circ$), controlling parameters $\varepsilon_1$ and $\varepsilon_2$, epoch count ($\mathcal{P}$), batch size ($b$), and device's labeled data ($\mathcal{D}^l_i$) and unlabeled data ($\mathcal{D}^u_i$)}
\KwOut{$M$ specialized models and a HFL global model}
  
\textbf{Initialization:} Define device clusters $\mathcal{I}=\{\{1,\dots,I_1\},\dots,\{1,\dots,I_J\}\}$, set $\theta_i \leftarrow 0~\forall i$, and $r=1$

\For{$r=1$ to $R$}
{
  \tcp{Aggregation at Edge Servers:}
  \For{$j=1$ to $J$}{
    \tcp{Edge server pre-processing:}
      {Select all available and active workers} \par
       {Edge server $j$ distributes $\theta_{r-1}$ to all workers}\par
      \If{$M>1$}{Edge server selects \par
      $\{m \in \mathcal{M}|~\underset{i\in m}{\max}\parallel \nabla_\theta F_i(\theta^*_j)\parallel<\varepsilon_2>0\}$}
\Else{Organize all individuals  to engage in the training}\par
      \textbf{Device tasks: Parallel participating}\par
      \For{$i=1$ to $|\mathcal{C}^j_{r}|$}{
          Receive $\theta_{r-1}$\\
           \tcp{Device's Pseudo Labeling}
          Perform local training:\par
          $\theta_i = \theta_{r-1} - \eta\sum_{t=1}^{\tau}\nabla 
          F_i(\theta_i{(t)})$\\
          Aggregate models, compute $F_j(\theta_r)$, and update $\theta_j$\par
          \If{Algorithm 1 is chosen}{
    Call Algorithm 1 for implementation
}
\ElseIf{Algorithm 2 is chosen}{
Call Algorithm 2 for implementation
}

         }

  } 
  \tcp{Aggregation at Cloud:}
  
      Aggregate edge and specialized models, updating global model\par
      \If{$\textbf{M}>2$}{perform similarity checks among specialized models}  
   Increment $r$
  }
{One HFL model and $M$ specialized models are developed}
\end{algorithm}
 The key difference from the scheme detailed in Scenario~\ref{Scenario6} is the timing and trigger for labeling and injecting unlabeled data. This scheme utilizes a stopping-based prediction time scheme, where the pseudo-labeling process is initiated once a stopping point is met for each cluster.
 \subsubsection{Weighted-averaging ensemble model with stopping-based prediction time and round-robin selection (EMSTRR)}
\label{Scenario8}
This scheme parallels the one described in Scenario  \ref{Scenario7}, where the only difference is the worker selection scheme. As discussed in previous schemes, we adopt the round-robin algorithm for worker selection after clusters reach their stopping points.

It is worth mentioning that the integrated schemes that use the best-performing specialized model are carefully designed to prevent overfitting, as explained below: First, the iterative clustering process allows for continual adjustment based on current data distributions. We develop specialized models tailored for specific tasks, each requiring a perfect model that fits its data distribution. Second, the specialized models reliably and effectively adapt, accurately labeling new data when it matches the patterns in the data used to develop these models. This shows their robustness in handling local data and generalizing unseen data well.

Algorithm \ref{CFL3} outlines the steps for the proposed approach, incorporating the prediction model, prediction time, and scheduling schemes in HWNs. Input parameters include worker count ($I$), initial parameters ($\theta_\circ$), controlling parameters ($\varepsilon_1$, $\varepsilon_2$), and labeled ($\mathcal{D}^l_i$) and unlabeled ($\mathcal{D}^u_i$) data. The algorithm starts by initializing workers based on availability and setting model parameters (lines 6-8). Workers perform local training, updating model parameters (line 15), followed by edge servers aggregating these updates (line 16). Edge servers either call Algorithm \ref{CFL1} or \ref{CFL2} based on the selected scheme (lines 18, 20) and compute cosine similarity to cluster workers (lines 2-9 in Algorithm \ref{CFL1}). Pseudo-labeling occurs during worker splitting (line 10 in Algorithm \ref{CFL1}) or when clusters reach a stopping point (line 1 in Algorithm \ref{CFL2}). Once pseudo-labels are injected (lines 12-13 in Algorithm \ref{CFL1} and lines 3-4 in Algorithm \ref{CFL2}), the dataset $\mathcal{D}^l_{i,new}$ is created for subsequent rounds. The aggregated models are then sent to the cloud for global updates (lines 22-23), and the process repeats until optimal results are achieved (line 25).
\section{Convergence Analysis}
\label{Conv}
In this section, we provide a detailed convergence analysis of our proposed approach. We start by defining the global objective function within the HFL framework as follows:
\begin{equation}
F(\theta) = \sum_{i=1}^{I} b_i F_i(\theta),
\end{equation}
where $F_i(\theta) $ is the local loss function for $i$-\textit{th} worker, $\theta$ is the model parameters, and $b_i = \frac{D^l_i}{D^l} $ is the weight associated with $i$-\textit{th} worker, based on the proportion of data $D^l_i$ they hold relative to the total data $D^l = \sum_{i=1}^{I} D^l_i $. The local loss function $F_i(\theta)$ is typically defined as:
\begin{equation}
\small
F_i(\theta)=\frac{1}{D^l_i} \sum_{s=1}^{ \mathcal{D}^l_i} \mathcal{L}(x^{(i)}_{s,d},y^{(i)}_s; \theta),
\end{equation}
where $\mathcal{L}(x^{(i)}_{s,d},y^{(i)}_s; \theta) $ is the loss for a single data point $(x^{(i)}_{s,d},y^{(i)}_s) $ in the $i$-\textit{th} worker's dataset.
In our proposed approach, workers are clustered according to their data distributions to effectively address the non-IID nature of the data. Let there be $M$ clusters, each $m$-\textit{th }cluster containing $I_m$ workers. The objective specific to each cluster is defined as follows:
\begin{equation}
F_m(\theta) = \frac{1}{I_m} \sum_{i \in g_m} F_i(\theta).
\end{equation}
Accordingly, the global objective function can then be rewritten as:
\begin{equation}
\small
F(\theta) = \sum_{m=1}^{M} \frac{I_m}{I} F_m(\theta) = \sum_{m=1}^{M} \frac{I_m}{I} \left( \frac{1}{I_m} \sum_{i \in g_m} F_i(\theta) \right).
\end{equation}
This formulation considers the clustering of workers and ensures that the global objective accurately reflects the data distribution across all clusters. In the semi-supervised setting, where labeled data is scarce, the adaptive labeling mechanisms (i.e., the best-performing specialized model and the weighted-averaging ensemble model schemes) modify the local loss function. The resulting changed local loss function is expressed as follows:
\begin{equation}
\small
\tilde{F}_i(\theta) = F_i(\theta) + \nu~\mathrm{Reg}_i(\theta),
\end{equation}
where $\mathrm{Reg}_i(\theta)$ is a regularization term representing the correction due to adaptive labeling, and $\nu $ is a hyperparameter controlling the influence of this regularization. The regularization term $\mathrm{Reg}_i(\theta)$ can be modeled as:
\begin{equation}
\small
\mathrm{Reg}_i(\theta)= \frac{1}{D^u_i} \sum_{z=1}^{D^u_i} \left( \mathcal{L}(x^{(i)}_{z,d},\hat{y}^{(i)}_z)-\mathcal{L}(x^{(i)}_{z,d},{y}^{(i)}_z) \right),
\end{equation}
where $\hat{y}^{(i)}_z $ represents the predicted label for an unlabeled data point $x^{(i)}_{z,d} $, and $D^u_i $ is the number of unlabeled data points for $i$-\textit{th} worker.
Consequently, the revised global objective function is expressed as:\begin{equation}
\small
\tilde{F}(\theta) = \sum_{m=1}^{M} \frac{I_m}{I} \left( \frac{1}{I_m} \sum_{i \in g_m} \tilde{F}_i(\theta) \right).
\end{equation}
Expanding this, we obtain:
\begin{equation}
\small
\tilde{F}(\theta) = F(\theta) + \vartheta \sum_{m=1}^{M} \frac{I_m}{I} \left( \frac{1}{I_m} \sum_{i \in g_m} \mathrm{Reg}_i(\theta) \right).
\end{equation}
To analyze convergence, we make the following assumptions, as in \cite{sattler2020clustered,10172335,8387465}:
\begin{itemize}
    \item \textbf{Lipschitz Continuity}: Each local loss function $F_i(\theta) $ is $L $-smooth, meaning:
    \begin{equation}
    \small
    \|\nabla F_i(\theta) - \nabla F_i(\theta')\| \leq L \|\theta - \theta'\|,
    \end{equation}
    which implies that the gradient of the local loss function does not change rapidly.
    \item \textbf{Strong Convexity}: The global objective $F(\theta) $ is $\mu $-strongly convex:
    \begin{equation}
    \small
    F(\theta) \geq F(\theta^*) + \frac{\mu}{2}\|\theta - \theta^*\|^2,
    \end{equation}
    where $\theta^* $ is the optimal solution. This ensures that the objective function has a unique minimum.
    \item \textbf{Bounded Variance}: The variance of the stochastic gradients is bounded by $\sigma^2 $:
    \begin{equation}
    \small
    \mathbb{E}[\|\nabla F_i(\theta) - \nabla F(\theta)\|^2] \leq \sigma^2.
    \end{equation}
\end{itemize}
The CFL algorithm updates the global model based on the aggregated gradients from each cluster. The update rule at round $r$ is:
\begin{equation}
\small
\theta_{r+1} = \theta_r - \eta_r \sum_{m=1}^{M} \frac{I_m}{I} \nabla \tilde{F}_m(\theta_r),
\end{equation}
where $\eta_r $ is the learning rate. Expanding the gradient of the modified objective function, we obtain:
\begin{equation}
\footnotesize
\nabla \tilde{F}_m(\theta) = \frac{1}{I_m} \sum_{i \in g_m} \nabla \tilde{F}_i(\theta)= \frac{1}{I_m} \sum_{i \in g_m} \left( \nabla F_i(\theta)+\lambda \nabla \mathrm{Reg}_i(\theta) \right).
\end{equation}
Thus, the global update is given by:
\begin{equation}
\footnotesize
\theta_{r+1} = \theta_r - \eta_r \left( \nabla F(\theta^r) + \nu \sum_{m=1}^{M} \frac{I_m}{I} \left( \frac{1}{I_m} \sum_{i \in g_m} \nabla \mathrm{Reg}_i(\theta_r) \right) \right).
\end{equation}
To analyze the convergence rate, we examine the expected reduction in the global objective function at each round. For a smooth and strongly convex function, this expected decrease is expressed as:
\begin{equation}
\footnotesize
\mathbb{E}[F(\theta_{r+1}) - F(\theta^*)] \leq \left(1 - \frac{2\mu\eta_r}{1 + \mu \eta_r}\right) \mathbb{E}[F(\theta_r) - F(\theta^*)] + \mathcal{O}(\eta_r^2 \sigma^2),
\end{equation}
where $\sigma^2$ represents the variance introduced by the stochastic nature of the updates. As previously mentioned, the adaptive labeling mechanism, including the prediction model and time schemes, introduces both bias and variance into the gradient estimates. The bias is represented by the term:
\begin{equation}
\small
\nu \sum_{m=1}^{M} \frac{I_m}{I} \left( \frac{1}{I_m} \sum_{i \in g_m} \nabla \mathrm{Reg}_i(\theta) \right),
\end{equation}
which is controlled by the regularization parameter $\nu $. The variance introduced by the adaptive labeling process can be expressed as:
\begin{equation}
\text{Var}(\nu \mathrm{Reg}_i(\theta)) = \nu^2 \text{Var}(\mathrm{Reg}_i(\theta)).
\end{equation}
This variance arises from the uncertainty in the predicted labels $\hat{y}^{(i)}_z$ and the heterogeneity of data distributions across clusters. Ideally, the bias introduced by the labeling process should be reduced as the model improves over time. Letting $ \mathrm{Reg}_i(\theta)$ is the difference between the true and predicted labels, we can bound the bias as follows:
\begin{equation}
\|\nu \nabla  \mathrm{Reg}_i(\theta)\| \leq \nu \|\nabla  \mathrm{Reg}_i(\theta)\| \leq \nu \rho,
\end{equation}
where $\rho $ is a bound on the regularization term. The choice of $\nu $ is crucial here: if $\nu $ is too large, the bias could dominate and slow down convergence; if $\nu $ is too small, the regularization effect might be negligible.
 The variance $\text{Var}(\mathrm{Reg}_i(\theta))$ can be bounded by considering the variability in the labeling process across different clusters:
\begin{equation}
\text{Var}(\mathrm{Reg}_i(\theta)) \leq \frac{\sigma_R^2}{I_m},
\end{equation}
where $\sigma_R^2$ is the maximum variance of labeling errors across all clusters. As the number of workers $I_m$ within each cluster increases, this variance decreases, facilitating convergence. Moreover, as the accuracy of predicted labels improves over time, the variance term $\sigma_R^2$ is expected to decrease, further promoting convergence.

Combining the results from the bias and variance analysis, the convergence rate can be approximated as:

\begin{footnotesize}
  \begin{align}
\mathbb{E}[F(\theta_{r+1}) - F(\theta^*)] \leq& \left(1 - \frac{2\mu\eta_r}{1 + \mu \eta_r}\right) \mathbb{E}[F(\theta_r) - F(\theta^*)] +\nonumber\\
&\mathcal{O}\left(\eta_r^2 \left(\sigma^2 + \vartheta^2 \sigma_R^2 \right)\right).
\end{align}  
\end{footnotesize}
\section{Performance Evaluation}
\label{results}
In this section, we evaluate the performance of the CFSL framework compared to baselines, concentrating on testing accuracy, labeling accuracy, and energy consumption metrics.
\subsection{Experimental Setup}
\textbf{Experimental Parameters:} The system's bandwidth is allocated at 10 MHz. The channel gain (${h}_i$) for each worker is estimated based on a path loss model ($PL=g_\circ (\frac{d_\circ}{d})^\upsilon$), where the path loss exponent ($\upsilon$) is set to 4, the reference gain ($g_\circ$) is set at -35 dB, and the reference distance ($d_\circ$) is fixed at 2 meters. The total noise power of AWGN ($N_\circ$) is $10^{-8}$ watts. Workers operate within a computation frequency ($f_i$) range of 1 GHz to 9 GHz, and each worker is assigned a computational requirement of $\Psi=20$ cycles/sample. The transmission power ($P_i$) is restricted to a range between $P^{\min}_i=-10$ dBm and $P^{\max}_i=20$ dBm.\\ 
\textbf{Dataset Preprocessing:}
We evaluate the performance of the proposed approach through extensive simulations using FEMNIST and CIFAR-10 datasets~\cite{gowal2021improving,caldas2018leaf}. These datasets are crucial for applications in handwriting recognition and object detection within HFL environments. The CIFAR-10 dataset is used in autonomous driving technologies, playing a vital role in image analysis, especially for vehicles' operating safety and efficiency. This dataset consists of $60,000$ images in 32x32 color format. Meanwhile, the FEMNIST is essential in categorizing letters and digits (A-Z, a-z, and 0-9), offering a collection of $244,154$ images for training and $61,500$ images designated for evaluation.  The dataset is split into $80\%$ for training and $20\%$ for testing. Our proposed approach addresses the challenge of multi-class classification in highly non-IID HWN environments, where traditional methods, such as naive local labeling, are insufficient for effectively labeling unlabeled data. To simulate extreme non-IID conditions that closely resemble real-world HWNs, the proposed approach adopts a data distribution strategy by dividing the datasets into ${\varrho}$ segments, ensuring that each worker has access to only two classes from the larger pool of classes in FEMNIST (i.e., 62 classes) and CIFAR-10 (i.e., 10 classes). This guarantees that each worker is assigned a unique subset of data, ensuring that no single worker completely understands the entire dataset. In addition, the dataset assigned to each worker is also divided into a small portion of labeled data and a much larger portion of unlabeled data, with the percentage of labeled data set at $5\%$, $10\%$, or $15\%$. As shown later, our experimental results demonstrate that the proposed approach outperforms baselines, maintaining high accuracy and significantly enhancing performance by optimizing the labeling process and resource consumption.

For model architecture, we use a convolutional neural network (CNN) for the FEMNIST dataset, featuring two hidden layers, while deploying the VGG-19 architecture, which is a more complex deep neural network (DNN), to analyze the CIFAR-10 dataset. Both models utilize the ReLU to activate the hidden layers, with the Softmax function activated in the output layer to facilitate the learning tasks.  In this regard, we evaluate the model's adaptability by training it with a limited quantity of labeled data and later evaluating its prediction accuracy on different unlabeled, unseen samples to show the model's flexibility in adjusting to new data scenarios. The simulation's parameters are configured to execute 200 communication rounds (R) involving 200 workers (I). The learning rate ($\eta$) is set at 0.01, and local training is performed over 10 epochs ($\mathcal{P}$) within each global round. Detailed simulation settings are concisely presented in Table~\ref{tab:setuppar}. \\  
\textbf{Benchmarks:} The performance of the proposed framework is evaluated and compared with different baseline techniques.
\begin{itemize}
   \item The CFL technique leverages only the labeled dataset for training. This approach involves intelligence scheduling (i.e., greedy and round-robin selections).
   \item The CFL technique with SSL uses prediction models and time schemes but with random selection for workers.
  \item The traditional HFL with SSL (HFSL), where only one global model is used for pseudo-labeling tasks under the non-IID data setting and variation of wireless channels.
\end{itemize}
\subsection{Results and Discussion}
 \subsubsection{The Significance of Applying the CFSL}
\begin{figure}[t]
    \centering
    \includegraphics[width=0.70\linewidth]{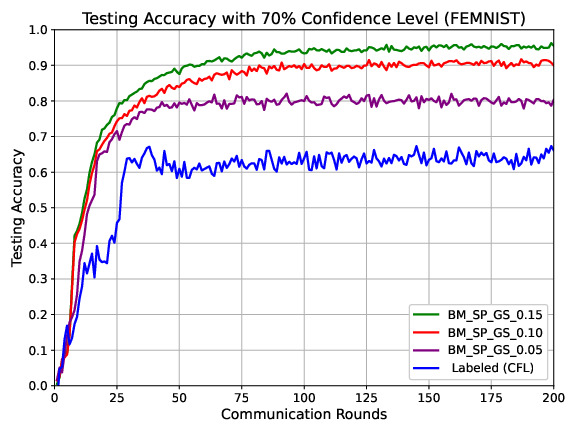}    \caption{Testing accuracy for the proposed approach compared to labeled CFL (without SSL) at $\Phi=0.70$.}
    \label{Label_VS_semi_1}
\end{figure}

We evaluate the proposed approach to illustrate its importance compared with two baselines: First, labeled CFL approach using intelligent worker selection strategies, and second, CFL with SSL using a random selection approach. We select the BMSPGS scenario as an example to show its superiority against labeled CFL with the greedy selection algorithm. Fig. \ref{Label_VS_semi_1} details the testing accuracy performance for both the proposed and baselines. Here, $\Phi$ is set to  $0.70$, and the ratio of labeled data to the total data is varied (i.e., $0.05, 0.10$, and $0.15$) to evaluate the impact of different portions of labeled data. One can observe that the proposed approach greatly outperforms the baseline in all labeled data proportions. This stems from leveraging unlabeled data by predicting its labels using the best-performing specialized scheme and injecting these pseudo-labeled samples into the training data for further improvement, enabling the system's overall efficacy. On the other hand, using only the labeled data limits the diversity and volume of data available for training, severely constraining the performance of the models. This arises because of the scarcity of labeled data used for the training, reducing the model's ability to generalize and capture complex patterns.
\begin{table}[t!]
\footnotesize
	\caption{Parameters of Simulation}
	\label{parameter}
	\centering
	\begin{tabular}{|c|p{4.5cm}|}%
		\hline
		\textbf{Parameter}&\textbf{Value}\\
		\hline
		No. of workers (I) & 200\\		\hline
        No. training rounds ($R$) & $200$\\		\hline

		 No. of edge servers (J) & 3\\		\hline

		 No. of epochs ($\mathcal{P}$) & 10\\		\hline

		 batch size (b) & 32\\		\hline

		 Learning rate ($\eta$) & 0.01\\		\hline

		System bandwidth ($B$) & 10 MHz\\		\hline

		 Transmission Power ($P_i$) & [$-10, 20$] dBm  \\		\hline

        AWGN's noise power ($N_\circ$) & $10^{-8}$ W \\		\hline

		 CPU Frequency ($f_i$) & [$1, 9$] GHz\\		\hline

        Cycles Per Sample $(\Psi)$ & 20 cycle/sample \\		\hline

		Confidence levels ($\Phi$) & $0.6, 0.7,0.80,$ and $0.90$\\		\hline
        Percentage of labeled data & $0.5,0.10,$ and $0.15$\\		
        \hline
	\end{tabular}
\label{tab:setuppar}
\end{table}
\begin{figure}[t]
    \centering
    \includegraphics[width=0.7\linewidth]{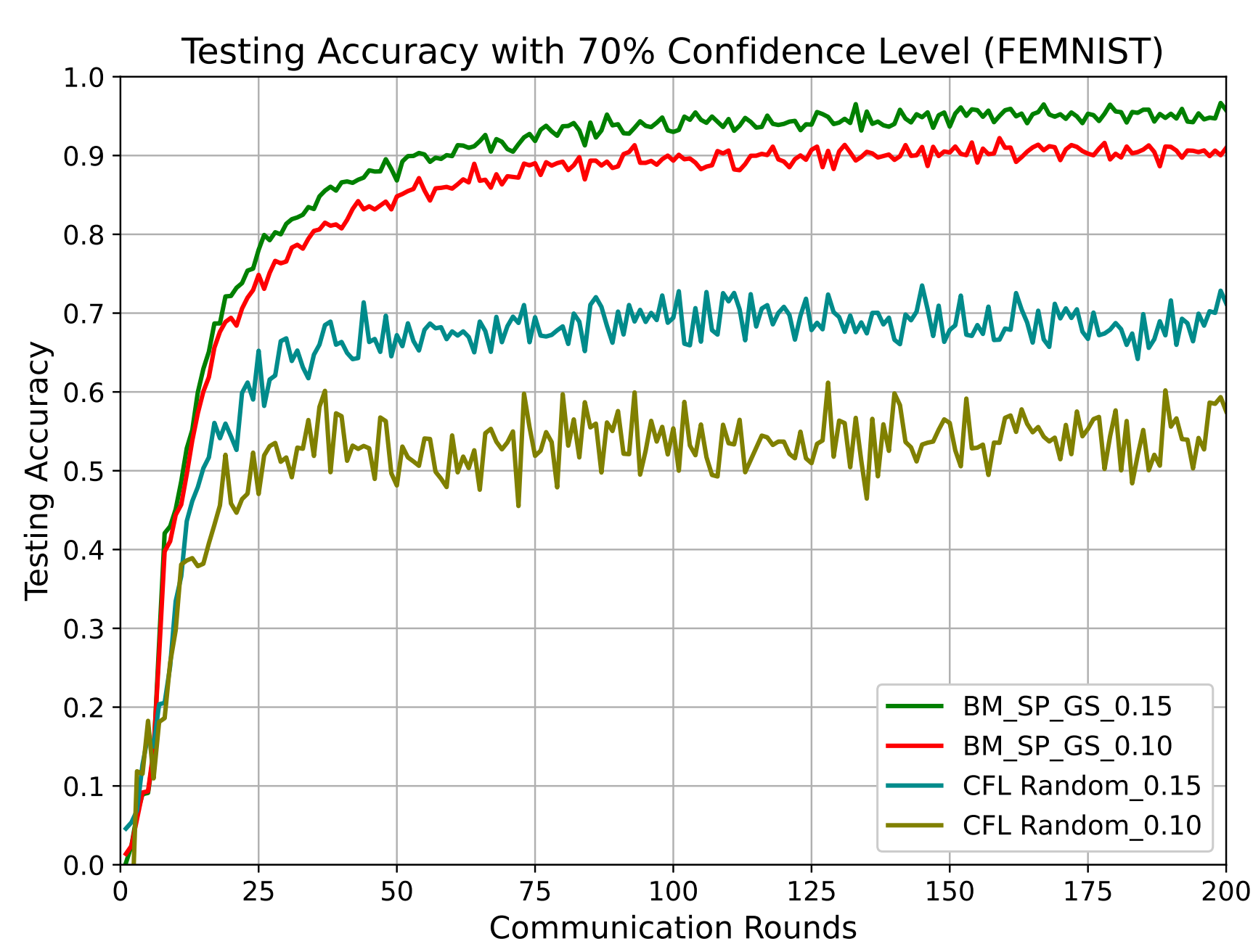}
    \caption{Testing accuracy for the proposed approach compared to CFL with SSL using random selection at $\Phi=0.70$.}
    \label{Schedule_VS_Random}
\end{figure}

Fig. \ref{Schedule_VS_Random} shows the performance of the proposed approach (BMSPGS) against the CFL with SSL that uses random worker selection to evaluate the significance of proposed worker selection in enhancing the overall performance. Here, $\Phi$ is also set at $0.70$, and the labeled-to-total data ratio is adjusted to $0.10$ and $0.15$ for both approaches. The results in this figure demonstrate that BMSPGS outperforms this baseline, attributed to its greedy selection strategy, which prioritizes workers with less latency and better resources, ensuring optimal resource utilization. In contrast, the baseline's random selection fails to adequately represent the data distribution, leading to potential biases and inconsistent model performance.
\subsubsection{Learning Performance Evaluation}
\begin{figure*}[t!]
    \centering  
\begin{subfigure}[b]{0.28\textwidth}
\centering 
\includegraphics[width=1\linewidth]{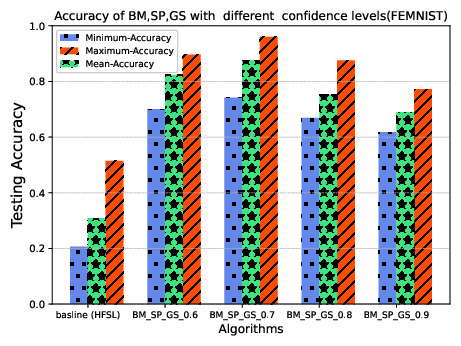}
\caption{BMSPGS.}
\label{BM_SP_GS}
\end{subfigure}
\begin{subfigure}[b]{0.28\textwidth}
\centering 
\includegraphics[width=1\linewidth]{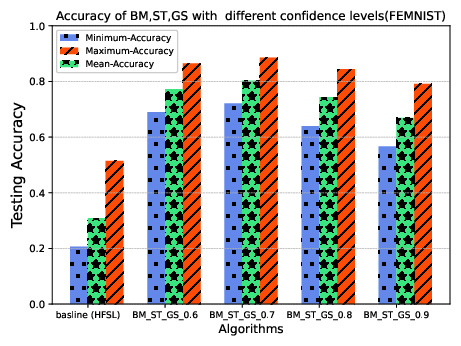}
    \caption{BMSTGS. }
    \label{BM_ST_GS}
\end{subfigure}
\begin{subfigure}[b]{0.28\textwidth}
\centering 
\includegraphics[width=1\linewidth]{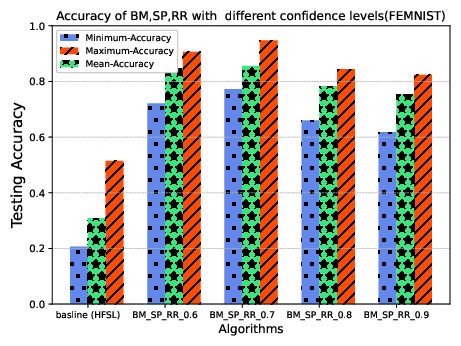}
\caption{BMSPRR.}
\label{BM_SP_RR}
\end{subfigure}
\begin{subfigure}[b]{0.28\textwidth}
\centering 
\includegraphics[width=1\linewidth]{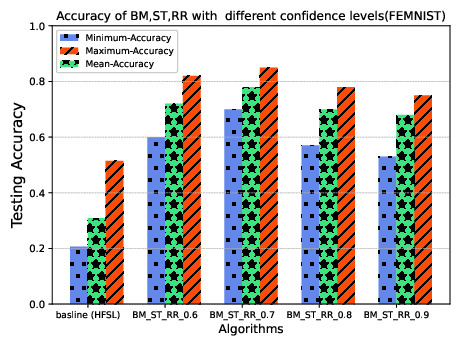}
    \caption{BMSTRR.}
    \label{BM_ST_RR}
\end{subfigure}
\begin{subfigure}[b]{0.28\textwidth}
\centering 
\includegraphics[width=1\linewidth]{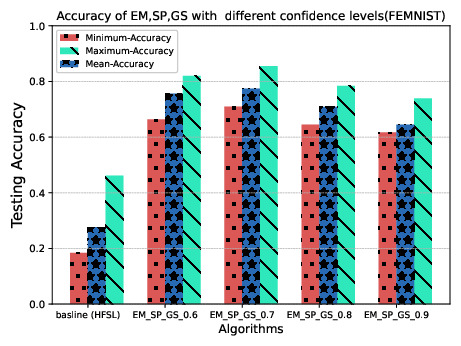}
    \caption{EMSPGS.}
    \label{EM_SP_GS}
\end{subfigure}
\begin{subfigure}[b]{0.28\textwidth}
\centering 
\includegraphics[width=1\linewidth]{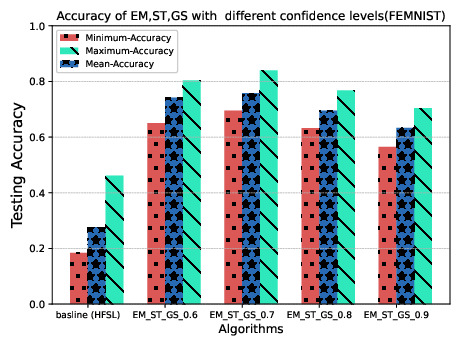}
    \caption{EMSTRR.}
    \label{EM_ST_GS}
\end{subfigure}
\caption{Minimum, average, and maximum accuracies for different scenarios using FEMNIST.}
\label{ACC_FEMNIST}
\end{figure*}

We evaluate proposed and baseline (i.e., HFSL approach) algorithms using the FEMNIST dataset. Fig. \ref{ACC_FEMNIST} shows testing accuracy performance, including the minimum, mean, and maximum accuracies. Note that Figs. \ref{BM_SP_GS} and \ref{BM_ST_GS} demonstrate the performance of the scenarios of best-performing specialized models with greedy selection and two different prediction time schemes: split-based and stopping-based, respectively, compared to baselines. One can notice that the proposed approach in both scenarios significantly outperforms the baselines.  For instance, the BMSPGS scenario at $\Phi=0.9$ demonstrates a notable $123.33\%$ increase in mean accuracy over baselines. This arises since our approach organizes workers according to their data distributions, allowing each worker to attain a specialized model finely tuned to their data distribution.

Similarly, the performance of scenarios of the best-performing specialized model with round-robin worker selection using the split-based and stopping-based prediction times is shown in Figs. \ref{BM_SP_RR} and \ref{BM_ST_RR}. In these scenarios, we have adopted a round-robin scheme to ensure an equitable selection of workers throughout the training. As shown in both figures, the proposed approach surpasses baselines with different $\Phi$ values (i.e., $0.6, 0.7, 0.8$, and $0.9)$. This arises from the effective clustering mechanism the CFSL follows to split workers into clusters, each with a specialized model, leading to more accurate labeling in a shorter time. Overall, all proposed scenarios in Figs. \ref{BM_SP_GS} to \ref{BM_ST_RR} significantly exceed minimum, mean, and maximum accuracy baselines with different confidence thresholds.

We also implemented the weighted-averaging ensemble model scheme to assess the testing accuracy in various scenarios, comparing them with the baselines. Figs. \ref{EM_SP_GS} and \ref{EM_ST_GS} illustrate the proposed approach's performance using the EMSPGS and EMSTRR scenarios. These scenarios not only suppress baselines in testing accuracy but also enable more robust and targeted learning environments. Precisely, it combines the specialized models into one predictive model that predicts labels for the unlabeled data, ensuring robustness and adaptability across different workers.
\begin{figure}[t!]
\centering
  \includegraphics[width=0.550\linewidth]{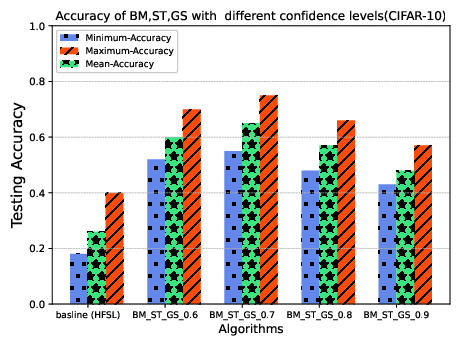}
\caption{ Testing accuracy for the BMSPGS (CIFAR-10).}
\label{CIFAR-10_1}
\end{figure}
 \begin{figure}[t!]
\centering
  \includegraphics[width=0.550\linewidth]{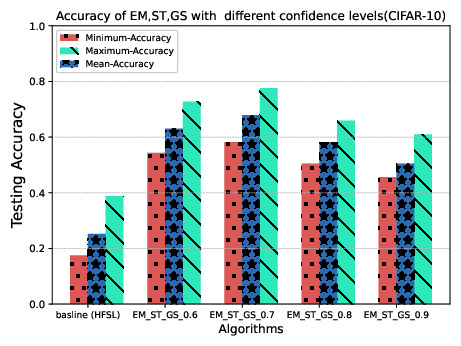}
\caption{  Testing accuracy for the EMSTGS (CIFAR-10).}
\label{CIFAR-10_2}
\end{figure}
 \begin{figure}[t!]
  \centering
 \begin{subfigure}[b]{0.4\textwidth}
 \centering
    \includegraphics[width=0.80\linewidth]{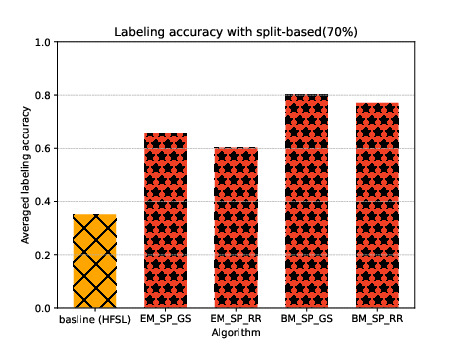}
    \caption{split-based prediction time. }
     \label{LA_BM_vs_EM}
\end{subfigure}
\begin{subfigure}[b]{0.4\textwidth}
\centering
    \includegraphics[width=.80\linewidth]{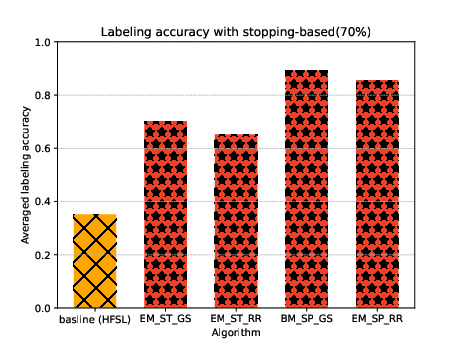}
    \caption{stopping-based prediction time.}
    \label{LA_BM_vs_EM_ST}
\end{subfigure}
\caption{Labeling accuracy for two prediction time schemes at ($\Phi=0.70$) using the FEMNIST dataset.}
     \label{labeling_acc}
\end{figure}
 \begin{figure*}[t!]
    \centering  
 \begin{subfigure}[b]{0.28\textwidth}
 \centering
    \includegraphics[width=1\linewidth]{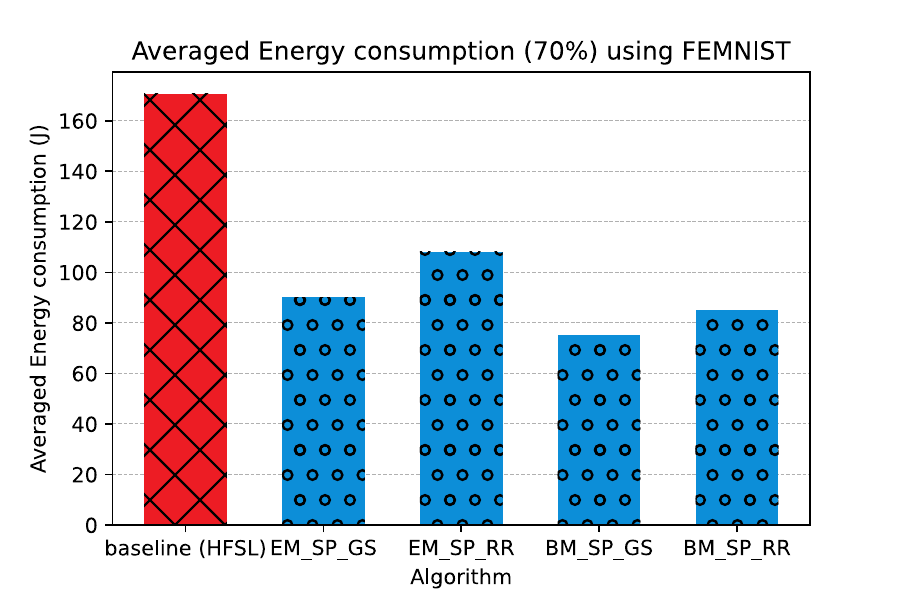}
    \caption{split-based prediction time using FEMNIST. }
    \label{Energy_split}
\end{subfigure}
\begin{subfigure}[b]{0.28\textwidth}
\centering
    \includegraphics[width=1\linewidth]{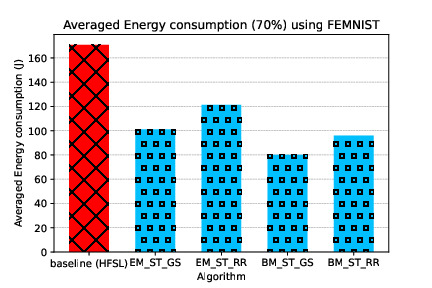}
    \caption{stopping-based prediction time using FEMNIST.}
    \label{Energy_round}
\end{subfigure}
\begin{subfigure}[b]{0.28\textwidth}
\centering 
\includegraphics[width=1\linewidth]{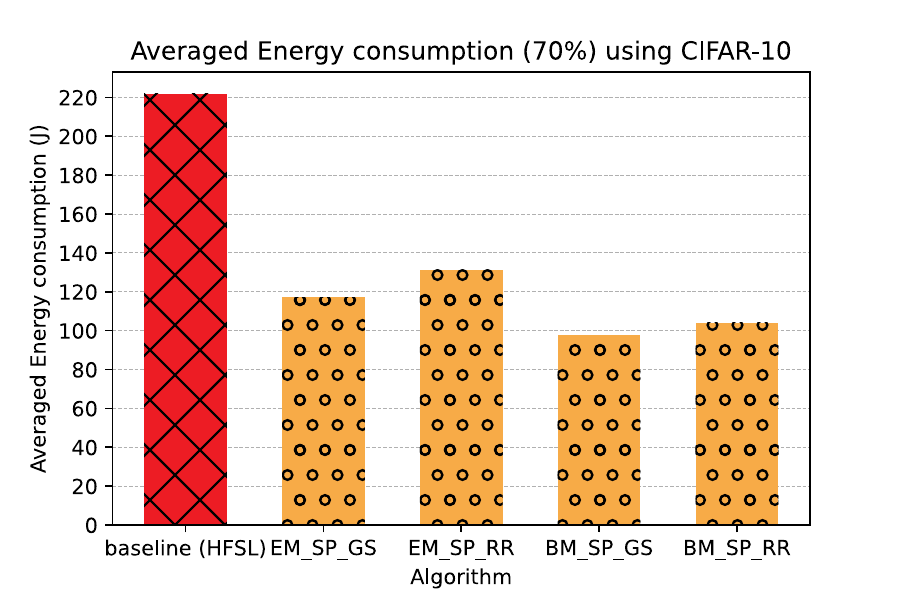}
    \caption{split-based prediction time using CIFAR-10.}
    \label{Energy_split_CIFAR10}
\end{subfigure}
\caption{Averaged energy consumption with split-based and stopping-based prediction time schemes at $\Phi=0.70$.}
\label{Tot_Energy}
\end{figure*}
\begin{figure}[t]
    \centering
    \includegraphics[width=0.550\linewidth]{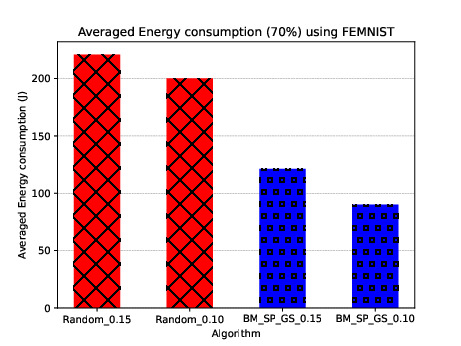}
    \caption{Averaged energy consumption of BMSPGS (Greedy) vs. CFL with SSL (Random) at $\Phi=0.70$.}
    \label{Energy_random}
\end{figure}

We further prove the CFSL's superiority over baselines in testing accuracy using the CIFAR-10 dataset and showcasing the performance of two selected scenarios (i.e., BMSTGS and the EMSTGS) within our framework to simulate more realistic scenarios. Fig. \ref{CIFAR-10_1} illustrates the testing accuracy of the BMSPGS scenario with different values of $\Phi$ ($0.60, 0.70, 0.80$, and $0.90$). For instance, our model's mean accuracy at $\Phi=0.90$ surpasses the baseline by $40.5\%$. This stems from the ability of specialized models to label unlabeled data at the appropriate time and best worker selection. Similarly, we evaluate the proposed approach using the EMSTGS scenario, as shown in Fig. \ref{CIFAR-10_2}. This scenario outperforms baselines since the weighted-averaging ensemble model combines the specialized models to perfectly fit the workers' data in their corresponding clusters. 
\subsubsection{Labeling Accuracy}
Evaluating our proposed approach against baselines using labeling accuracy reveals its superior capacity to label unlabeled, unseen, new data. Figs. \ref{LA_BM_vs_EM} and \ref{LA_BM_vs_EM_ST} exhibit the labeling accuracy of the proposed (i.e., with split-based and stopping-based label prediction, respectively) and baseline algorithms at $\Phi=0.70$ using the FEMNIST. Note that the proposed approach significantly improves labeling accuracy for all of our scenarios since we use specialized models to label unlabeled data, surpassing baseline performance. For instance, the BMSPRR scenario achieves a $71.26\%$ improvement in labeling accuracy over baselines, which stems from the specialized models' nuanced adaptation to the data distribution, achieving more accurate label prediction. In contrast, baselines develop only one global model that fails to capture all data patterns in non-IID data settings, leading to incorrect labeling.
\subsubsection{Energy Consumption Evaluation}
As shown in Fig. \ref{Tot_Energy}, the proposed approach is evaluated regarding the averaged energy consumption and compared with the baselines using the FEMNIST and CIFAR-10 datasets. Using the split-based prediction time scheme, Fig. \ref{Energy_split} shows the proposed approach's performance in the averaged energy consumption across different scenarios. Note that our scenarios consume less energy than the baselines during training and labeling at $\Phi=0.70$. For instance, the EMSPGS scenario reduces the averaged energy consumption by $47.22\%$, while the EMSPRR scenario reduces the averaged energy consumption by $36.67\%$. Also, the BMSPGS and BMSPRR scenarios demonstrate a significant reduction in energy consumption compared to baselines, achieving savings of $56.01\%$ and $50.27\%$, respectively. This stems from the CFSL's efficiency in developing specialized models that reach optimal solutions more swiftly and exhibit a better fit for diverse data distributions across edge networks in HWNs. This reduction is also attributed to the intelligent worker selection, either greedy or round-robin selection. Regarding the stopping-based prediction time scheme, the average energy consumption for the proposed approach when using the FEMNIST dataset is less than that in baselines, as shown in Fig. \ref{Energy_round}, which stems from the ability of the CFSL to label unlabeled data accurately compared to baselines.

To explore the CFSL's flexibility and adaptability across different applications, we analyzed its resource consumption using the CIFAR-10 dataset, as illustrated in Fig. \ref{Energy_split_CIFAR10}. This analysis emphasizes the proposed approach's energy efficiency compared to baselines. For instance, the CFSL with EMSPGS and EMSPRR scenarios achieves energy reductions of $47.18\%$ and $40.78\%$, respectively. Similarly, the BMSPGS and BMSPRR scenarios show energy savings of $55.98\%$ and $53\%$, respectively. This significant saving arises from the efficient labeling of unlabeled data, accurate labeling time, and the strategic selection of workers. Ultimately, our proposed approach not only reduces energy consumption across various datasets (i.e., FEMNIST and CIFAR-10) but also proves its effectiveness in HWN environments, proving its success in practical solutions attuned to real-world complexities.

The proposed approach (i.e., BMSPGS) is further evaluated against baselines (i.e., CFL with SSL but with random worker selection) in terms of the averaged energy consumption. Fig. \ref{Energy_random} shows that BMSPGS, using a greedy selection strategy, significantly reduces energy consumption across different labeled data percentages ($0.10$ and $0.15$) compared to baselines. This is attributed to the greedy algorithm's ability to prioritize workers with lower latency and better resources, minimizing energy costs. In contrast, the baseline's random selection fails to capture incongruent data distributions, delaying convergence and increasing energy consumption.  
\subsubsection{Lessons Learned}
\begin{itemize}
      \item The proposed approach significantly improves the model performance (i.e., testing and labeling accuracies) compared to baselines since specialized models are tailored to fit each worker's unique data distribution, ensuring a more targeted and efficient learning process.
       \item Labeling the unlabeled data using specialized models enhances model performance since they capture more details and accommodate diverse data patterns in HWNs.
        \item  Reducing the energy consumption for the proposed approach due to the efficient pseudo-labeling techniques and the flexible scheduling schemes (i.e., greedy and round-robin selections). 
        \item Overall, the proposed approach outperforms baselines by delivering superior testing and labeling accuracies while significantly reducing time and energy costs.
\end{itemize} 

\section{Conclusion}
\label{conclusion}
In this work, we have proposed a novel CFSL framework for HWNs, integrating three main schemes: prediction model, prediction time, and worker selection. The proposed approach focused on labeling unlabeled data using specialized models developed from the clustering mechanism while addressing data and system heterogeneities and resource constraints. We have proposed two prediction model schemes, the best-performing specialized model and the weighted-averaging ensemble model, achieving perfect labeling for unlabeled data. We also have introduced two techniques for optimal prediction timing: split-based and stopping-based, and two worker selection strategies, greedy and round-robin methods, for efficient scheduling. Given the scarcity of labeled data in non-IID data and limited resources, we have formulated the optimization problem to attain the best prediction model in the optimal time using intelligent work scheduling.  We have introduced a convergence analysis of the proposed approach to validate its efficiency. Our results show the CFSL approach's superiority in enhancing model performance while reducing resource consumption compared to baseline techniques. For future work, we aim to address learning challenges in dynamic environments with fluctuating data to improve our framework's effectiveness and adaptability.
\section*{Acknowledgement}
{This publication was made possible by TUBITAK-QNRF joint Funding Program (Tubitak-QNRF 4th Cycle grant \# AICC04-0812-210017) from the Qatar National Research Fund (a member of Qatar Foundation). The findings herein reflect the work, and are solely the responsibility of the authors.}
\bibliographystyle{IEEEtran}
\bibliography{Ref}
\begin{IEEEbiography}[{\includegraphics[width=1.1in,height=1.22in,clip]{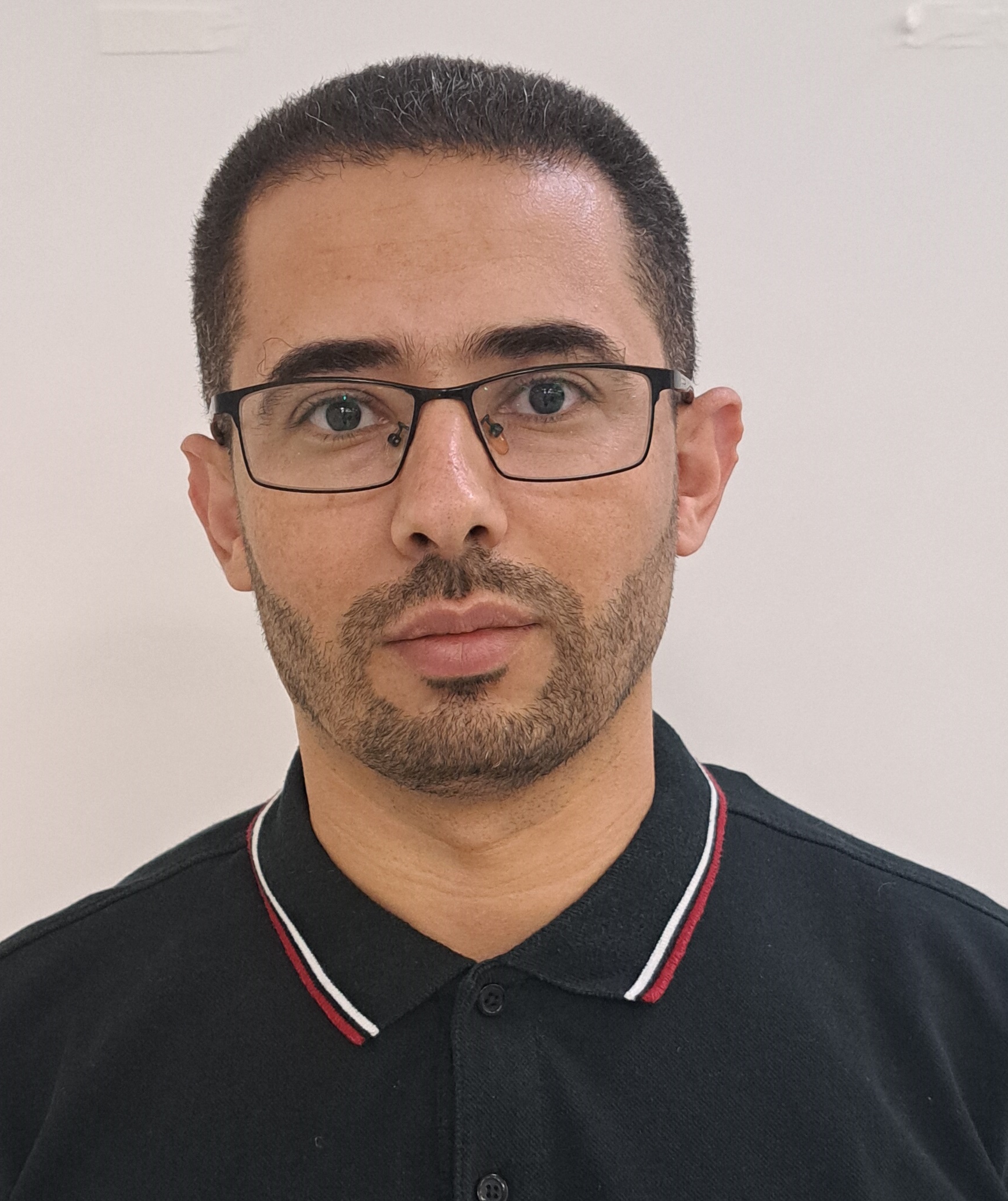}}]  
{Moqbel Hamood} received his B.Sc. degree in Electrical Engineering from Mutah University, Jordan, in 2012 and his M.Sc. degree in Wireless Communications from Jordan University of Science and Technology (JUST), Irbid, Jordan, in 2018. He is currently pursuing his Ph.D. at the Smart Communication Networks \& Systems Lab at Hamad Bin Khalifa University, Doha, Qatar. His research interests include federated learning over wireless networks.
\end{IEEEbiography}
\begin{IEEEbiography}[{\includegraphics[width=1.1in,height=1.22in,clip]{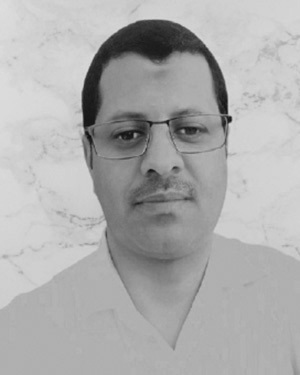}}]  
{Abdullatif Albaseer (Member, IEEE)} received his M.Sc. degree (with honors) in computer networks from King Fahd University of Petroleum and Minerals, Dhahran, Saudi Arabia, in 2017, and his Ph.D. degree in computer science and engineering from Hamad Bin Khalifa University, Doha, Qatar, in 2022. He is currently a Postdoctoral Research Fellow with the Smart Cities and IoT Lab at Hamad Bin Khalifa University. He was a Visiting Scholar at Texas A \& M University in Qatar from 2022 to 2023.
Dr. Albaseer has authored and co-authored over thirty journal and conference papers, primarily in IEEE Transactions. He also holds six US patents in the area of wireless network edge technologies. His current research interests include AI for Networking, AI for Cybersecurity, and the application of Large Language Models (LLMs) in both fields.
Dr. Albaseer has served as a chair and organizing committee member for international conferences. He is a reviewer for numerous prestigious IEEE journals and conferences. He has also presented at various international conferences and participated in numerous academic and professional development activities.
\end{IEEEbiography}
\begin{IEEEbiography}[{\includegraphics[width=1.1in,height=1.22in,clip]{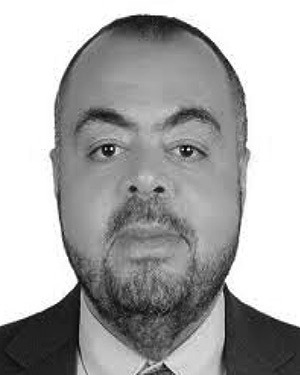}}]  
{Mohamed Abdallah (Senior Member, IEEE)} received the B.Sc. degree from Cairo University, in 1996, and the M.Sc. and Ph.D. degrees from the University of Maryland at College Park in 2001 and 2006, respectively. From 2006 to 2016, he held academic and research positions with Cairo University and Texas A\&M University at Qatar. He is currently a Founding Faculty Member with the rank of an Associate Professor with the College of Science and Engineering, Hamad Bin Khalifa University. He has published more than 150 journals and conferences and four book chapters, and co-invented four patents. His current research interests include wireless networks, wireless security, smart grids, optical wireless communication, and blockchain applications for emerging networks. He was a recipient of the Research Fellow Excellence Award at Texas A\&M University at Qatar in 2016, the Best Paper Award in multiple IEEE Conferences, including the IEEE BlackSeaCom 2019, the IEEE First Workshop on Smart Grid and Renewable Energym in 2015, and the Nortel Networks Industrial Fellowship for five consecutive years from 1999 to 2003. His professional activities include an Associate Editor of the IEEE Transactions on Communications and the IEEE Open Access Journal of Communications, the Track Co-Chair of the IEEE VTC Fall 2019 Conference, the Technical Program Chair of the 10th International Conference on Cognitive Radio Oriented Wireless Networks, and a Technical Program Committee Member of several major IEEE conferences.
\end{IEEEbiography}
\begin{IEEEbiography}[{\includegraphics[width=1.1in,height=1.22in,clip]{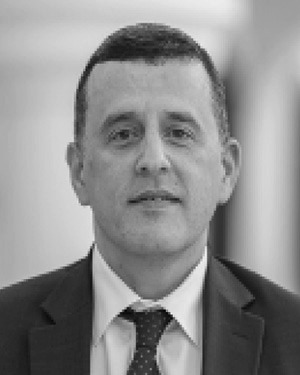}}]  
{Ala Al-Fuqaha (Senior Member, IEEE)} received the Ph.D. degree in computer engineering and networking from the University of Missouri-Kansas City, Kansas City, MO, USA, in 2004. He is currently a Professor with Hamad Bin Khalifa University. His research interests include the use of machine learning in general and deep learning in particular in support of the data-driven and self-driven management of large-scale deployments of IoT and smart city infrastructure and services, wireless vehicular networks (VANETs), cooperation, and spectrum access etiquette in cognitive radio networks, and management and planning of software defined networks. He serves on editorial boards of multiple journals, including IEEE Communications Letter and IEEE Network Magazine. He also served as the Chair, the Co-Chair, and a Technical Program Committee Member of multiple international conferences, including IEEE VTC, IEEE Globecom, IEEE ICC, and IWCMC. He is an ABET Program Evaluator.
\end{IEEEbiography}
\end{document}